\documentclass[10pt,conference,letterpaper]{IEEEtran}
\usepackage{graphicx}
\usepackage{url}
\usepackage{todonotes}
\usepackage{tabularx}
\usepackage{color}
\definecolor{StringColor}{rgb}{0.1,0.55,0.1}
\usepackage{listings}
\usepackage{paralist}
\usepackage{amsmath,amssymb}
\usepackage{balance}

\begin{document}

\title{DualTable: A Hybrid Storage Model for Update Optimization in Hive}

\author{%
Songlin Hu{\small $^{\#,1}$}, Wantao Liu{\small $^{\#,2}$}, Tilmann Rabl{\small $^{\dagger,3}$}, Shuo Huang{\small $^{\#,4}$}, \\

 Ying Liang{\small $^{\#,5}$}, Zheng Xiao{\small $^{\S,6}$}, Hans-Arno Jacobsen{\small $^{\dagger,7}$}, Xubin Pei{\small $^{\ddagger,8}$}, Jiye Wang{\small $^{\ast,9}$}\\%
\selectfont\rmfamily\itshape
$^{\#}$\,Institute of Computing Technology, Chinese Academy of Sciences, China\\
$^{\ddagger}$\,Zhejiang Electric Power Corporation, China\\
$^{\S}$\,State Grid Electricity Science Research Institute, China\\
\selectfont\rmfamily\itshape
$^{\dagger}$\,Middleware Systems Research Group, University of Toronto, Canada\\
$^{\ast}$\,Dept. of Information Technology, State Grid Corporation of China, China\\
\{husonglin$^1$,liuwantao$^2$,liangy$^5$\}@ict.ac.cn, tilmann.rabl@utoronto.ca$^3$, hshuocn@gmail.com$^4$,\\
 xiaozheng@sgcc.com.cn$^6$, jacobsen@eecg.toronto.edu$^7$, pxb@zj.sgcc.com.cn$^8$, jiyewang@sgcc.com.cn$^9$\\
}

\maketitle


\begin{abstract}
Hive is the most mature and prevalent data warehouse tool providing SQL-like
interface in the Hadoop ecosystem. It is successfully used in many Internet
companies and shows its value for big data processing in traditional
industries. However, enterprise big data processing systems as in Smart Grid
applications usually require complicated business logics and involve
many data manipulation operations like updates and deletes. Hive
cannot offer sufficient support for these while preserving high query performance. Hive
using the Hadoop Distributed File System (HDFS) for storage cannot implement
data manipulation efficiently and Hive on HBase suffers from poor query
performance even though it can support faster data manipulation.
There is a project based on Hive issue Hive-5317 to support update operations,
but it has not been finished in Hive's latest version. Since this ACID compliant
extension adopts same data storage format on HDFS, the update performance problem
is not solved.

In this paper, we propose a
hybrid storage model called DualTable, which combines the efficient streaming
reads of HDFS and the random write capability of HBase. Hive on DualTable provides
better data manipulation support and preserves query performance at the same time.
Experiments on a TPC-H data set and on a real smart grid data set show that
Hive on DualTable is up to 10 times faster than Hive when executing update and
delete operations.
\end{abstract}
%
%
%

\section{Introduction}
The Hadoop ecosytem is the quasi-standard for big data analytic applications.
It provides HDFS as a new file
system treating files as consistency unit, which makes it possible to
significantly improve batch data reading and writing \cite{Borthakur2007}.
Hive is a data warehouse system based on Hadoop for batch analytic query processing
\cite{Thusoo2009}. It has become very popular in Internet companies.

The success and ease of deployment of Hive attracts attention from traditional industries,
especially when facing large data processing challenges. Smart Grid applications, as
typical use cases, have to deal with enormous amounts of data generated by millions of smart
meters. For instance, the Zhejiang Grid, a province-level company in China,
currently owns about 17 million deployed smart meters, which will be increased to
23 million within 2 years.
According to the China State Grid standard, each of these meters needs to record data
and send it to the data center 96 times per day. The system has to support efficient
querying, processing and sharing of these enormous amounts of data,
which add up to 60 billion measurements per month only on province level.
The whole system needs to support user electricity consumption computing,  district
line loss calculating, statistics of data acquisition rates, terminal traffic statistics,
exception handling, fraud detection and analysis, and more amounting to around 100,000 lines
of SQL stored procedures in total.

As requested by the State Grid, the computing task must be finished from 1am to 7am
every day, or it will affect the business operations in working hours. In fact, the
processing cost of these stored procedures is so high that current solutions
based on relational database management systems (RDBMS) deployed on a high
performance 2*98 core cluster and an advanced storage system can hardly
complete the analysis in time. Even with the current number of smart meters and a
comparably low frequency of data collection of a single measurement per day,
the performance of current commercial
solutions is not acceptable after careful system optimizations carried out by
professional database administrators and business experts. For instance,
due to sophisticated join operations on 5 tables that contain 60G data, around 1 billion data records in total,
the average processing time of the user electricity consumption is around 3 hours.
With increasing collection frequencies and a growing number of installed meters
the capacity of the current solution will be exceeded soon.
Considering the advantages of Hadoop and Hive, such as superior scalability,
fault tolerance, and low cost of deployment they were chosen for the Zhejiang Grid.
The use of Hive makes pure statistical applications in Zhejiang Grid more efficient. The performance of 
some statistical query executed in a Hive cluster is significantly better than that of current RDBMS cluster.

The main challenge in this use case is that current Hive lacks the capability of supporting
efficient data manipulation operations. 
Although HIVE-5317 aims at implementing insert, update, and delete in Hive with full
ACID support, it has not been released yet \cite{Hive-5317}.
Meanwhile, judging from its design document,
its main focus is on full ACID guarantee rather than performance optimization of update operation.

This makes it very difficult for
current RDBMS-based applications to be migrated to a Hadoop environment.
Traditional enterprises have to process
complicated business logic functions rather than only pure statistical applications. Many of the enterprise level data processing
applications are built using complex stored procedures. Besides
sophisticated analysis on huge data, they contain a high ratio of update and delete
operations. As shown in our analysis of the Zhejiang Grid smart electricity
consumption information system, a typical application, which can have more than 10,000
lines of stored procedure code, includes 70\% data manipulation operations \cite{Liu:2014}. Without full update support,
specifically missing UPDATE, DELETE and the proprietary MERGE INTO operations, Hive has to use
\emph{INSERT OVERWRITE} to rewrite huge HDFS files even if only 1\% of
the complete data set is modified. As a result, the lack of update support in Hive results in
huge I/O costs, which will cancel out all the performance benefits.

The weakness of the Hive data manipulation operations lies in its storage subsystem: HDFS or HBase. HDFS is
designed for a \emph{write once read many} scenario and is good at batch reading.
It treats a whole file as consistency unit without any support of random writes.
HBase provides record level consistency to support efficient random reads and writes at
the cost of batch reading efficiency. Choosing either one of these two as the
underlying storage will sacrifice the benefits of the other, resulting in severe side-effect when
facing complex workloads.

As described by the design document, the ongoing implementation of Hive-5317 proposes
an approach to support data manipulation by using a base table and several delta tables.
Unmodified data is stored in the base table, and each transaction creates a delta table.
The read operation retrieves a record from base table and \emph{merges} it with
corresponding records in delta tables to get the up-to-date data view. However, due to
the usage of same storage format, the performance problem is not solved in this approach.

To combine the benefits of file-level consistency and record-level
consistency, and thus to support high throughput batch reads and efficient random
writes in a unified way, DualTable, a hybrid storage model is proposed in this
paper. It enables efficient reads and random writes through integration of two different storage formats.
A cost model-based adaptive mechanism dynamically selects the most efficient
storage policy at run-time. The data consistency can transparently be
 maintained by our \emph{UNION READ} approach. The use of random read capability of HBase makes the \emph{UNION READ} efficient. With the support of
DualTable, update capability of Hive can be enhanced without losing its batch read
efficiency.

The Smart Grid use-case is
presented in Section \ref{sec:Motivaton}. 
We then give a detailed analysis of the weakness of data manipulation operations in Hive in Section
\ref{sec:limitation}. Section \ref{sec:dualtable} presents Dualtable. Section \ref{sec:costmodel}
discusses DualTable's cost model. The implementation and evaluation will be given
in Section \ref{sec:impl} and Section \ref{sec:eval} respectively. We will introduce related work on Hive optimization in Section \ref{sec:related}.
Finally, conclusions and future work are presented in Section \ref{sec:conclusion}.

\section{Smart Grid} \label{sec:Motivaton}

The smart electricity consumption information collection system is an very
important part of smart grid, which acts as a mediator between electricity
consumers and the grid.

\subsection{Smart Electricity Consumption Information Collection System} \label{sec:collection_system}



The smart electricity consumption information system makes it possible for
the energy provider to be aware of the
quasi-real-time electricity consumption and to improve its businesses such
as electricity supply and pricing policy through deeply analyzing
and utilizing the data it collects. Different from traditional collection systems,
which mostly focus on support of billing processes based on once-per-month
data collection, it collects data hundreds of times per day and serves
as an intelligent service for diverse applications in the life
cycle of marketing, production, and overhauling of the grid as well as
a data source for interactive user service.

\begin{figure}
  \centering
  \includegraphics[width=\columnwidth]{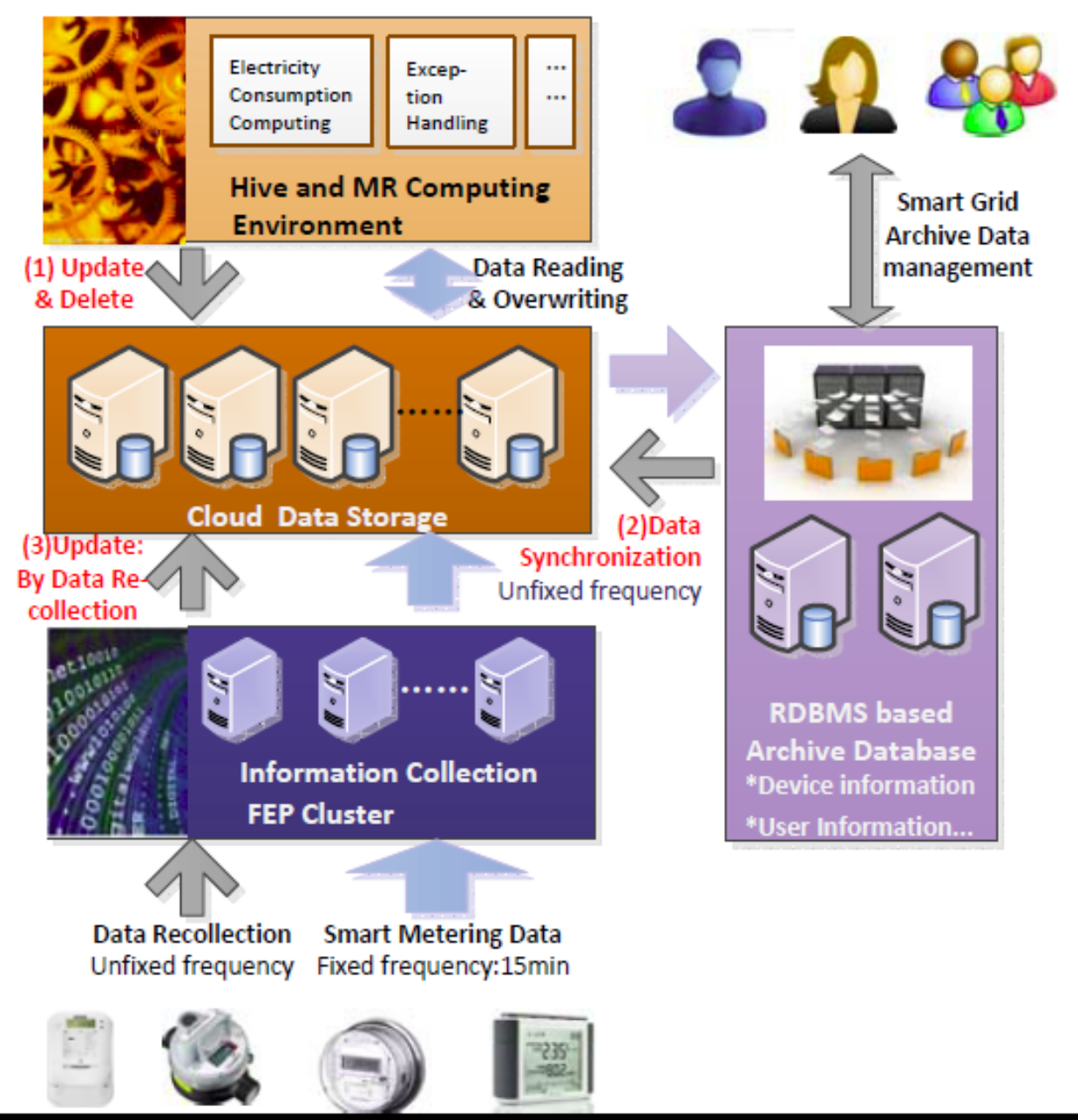}
  \caption{The Architecture and Data Flow of the Smart Electricity Consumption Information Collection System}
  \label{SystemArch}
\end{figure}

Considering its advantages, like cost-efficiency, fault tolerance, and
scalability, the Zhejiang Grid introduced Hive into its information
collection system and leveraged it as its big data processing platform.
The platform contains 5 subsystems as pictured in Figure \ref{SystemArch}: the communication system
that collects data from smart meters and sends them to the cloud after
encoding, the information collection cluster, i.e., front end PC (FEP), that receives the data
and does pre-processing like decoding, the cloud data storage system
that receives data from the FEP cluster and stores it, the Hive and
MapReduce environment that processes analytic procedures on the cloud
storage, and an RDBMS-based archive database that copes with daily data
management transactions on archive information of devices (smart metering
devices and inter-media devices), users, organizations, etc.

The data flow within the system is illustrated in Figure \ref{SystemArch}.
The collection system gets smart metering data at fixed frequency,
currently every 15 minutes. In cases of missing data or errors, the system
manager re-collects data from specified smart meters. The FEP Cluster usually appends
huge amount of the collected data to cloud data storage, thus, it needs a
very fast storage system to store the data. When recollection happens, it
needs to update the data set, which is marked as (1).

The archive database communicates with data managers and maintains data
according to their requests. To support grid data analysis, the archive
information involved will be copied to the cloud storage. And, the modified
data needs to be forwarded to the cloud storage via data synchronization marked as (2).

The computing environment executes all data processing procedures several
times per day and the results it generates are written back by the cloud storage
system to the RDBMS database for query and management. It reads data from the cloud
data storage and overwrites tables when needed. Moreover, as what has been implemented
in the RDBMS, it needs to update or delete only a small part of a table during data
processing. This is marked as (3).

From the perspective of the cloud storage system the computing environment has to cope
with data updating and deleting besides data appending and inserting, which is currently
supported by Hive system. In HiveQL, these operations must be implemented using current INSERT
OVERWRITE operation. Since a table in the collection system is very big, the OVERWRITING
operation will be very costly, which in turn heavily affects the efficiency, and sometimes
exhausts the resources and blocks the whole system.

\subsection{Hive Data Manipulation Limitations} \label{sec:limitation}


In real world enterprise data analysis use-cases, there is a high ratio of update
operations, as shown in Table \ref{tab:dml-ratio}. The following paragraphs will discuss the three update cases shown in Figure \ref{SystemArch} in detail respectively.
%

Frist, the SQL DML operations DELETE, UPDATE, and MERGE INTO, which updates existing records and
inserts new records, are extensively used in smart grid applications as illustrated in Figure \ref{SystemArch} (3).
For example, in the Zhejiang Grid data processing system, there are 5 important application scenarios:
\begin{inparaenum}[(i)]\item power line loss analysis, \item electricity consumption
statistics, \item data integrity ratio analysis, \item end point traffic statistics,
and \item exception handling. \end{inparaenum}  Each of these was implemented in stored
procedures in a traditional RDBMS, the total count of SQL code lines is more than 10,000 per application scenario. 
Each of the operations is executed more than 3 times per day.
Table \ref{tab:dml-ratio}
summarizes the amount of DML statements in the five core business scenarios. The table shows that DML operations
(UPDATE, DELETE, and MERGE) amount to at least 50\% in every scenario.
Note that Hive features efficient INSERT operations, which is why we
do not list INSERT in this table.

\begin{table}
\begin{center}
\begin{tabular}{l|lllll}
 Scenario & Total & Delete & Update & Merge & \% DML\\ \hline
1 & 133 & 15 & 52 & 15 & 62\\
2 & 75 & 25 & 20 & 9 & 72\\
3 & 174 & 27 & 97 & 13 & 79\\
4 & 12 & 3 & 3 & 0 & 50\\
5 & 41 & 3 & 23 & 0 & 63
\end{tabular}
\caption{Ratio of DML Operations in Grid Scenarios}
\label{tab:dml-ratio}

\end{center}
\end{table}



Due to its initial target use cases and limitation of HDFS, Hive lacks adequate support
for DML operations. Hive only supports \emph{complete overwrite} (INSERT OVERWRITE),
\emph{append} (INSERT INTO), and \emph{delete} (DROP) at table or partition level. Although
row-level UPDATE and DELETE operations can be transformed into equivalent INSERT OVERWRITE
statements, it is a tedious and error-prone process, let alone the complex logic correlations
and huge number of DML statements in an enterprise data analysis system.

To illustrate the challenges of transforming a data manipulation statement from SQL to
HiveQL, we show a typical UPDATE statement in Listing \ref{lst:RDBMSsql-update} and
its corresponding HiveQL translation in Listing \ref{lst:hive-update}
 from the electricity information collection system. The UPDATE
statement is part of the application scenario, which computes the total line loss of an
organization on a specific date from table tj\_tqxs\_r and changes the value of column QRYHS
in table tj\_tqxsqk\_r. As a comparison, in order to update only one column, Hive reads every
record and a total of 22 columns from table tj\_tqxsqk\_r, conducts a left outer join with
table tj\_tqxs\_r, and finally, writes back all 23 columns of every record into table
tj\_tqxsqk\_r using INSERT OVERWRITE. It is obvious that accessing the irrelevant
columns and records incurs high overhead.
Using INSERT OVERWRITE, the cost of a update operation is always proportional to total
amount of data instead of the amount of modified data. This leads to a significant performance
penalty for Hive if used as enterprise data analysis systems, which typically contain tables
with huge number of records and columns. Especially, when only a small portion of records
and columns are updated or deleted per operation.
In addition, the use of INSERT OVERWRITE is not as intuitive as the SQL's counterpart UPDATE.

\begin{lstlisting}[float=h, caption=SQL Update Statement, label=lst:RDBMSsql-update]
UPDATE tj_tqxsqk_r t
   SET t.QRYHS = (SELECT SUM(k.tqyhs)
       FROM tj_tqxs_r k
      WHERE t.rq = k.tjrq AND k.glfs = t.glfs
        AND k.zjfs = t.cjfs AND k.dwdm = t.dwdm
        AND k.sfqr = 1)
 WHERE t.rq = v_date;
\end{lstlisting}

\begin{lstlisting}[float=h, caption=Hive Update Statement, label=lst:hive-update]
INSERT OVERWRITE TABLE tj_tqxsqk_r
  SELECT t.dwdm,t.rq,t.jb,t.xslzctqs,t.xslcdtqs,
         t.xslwftqs,t.ztqs,t.xslzcyhs,t.zyhs,
         t.tqxsksl,t.glfs,t.cjfs,t.qfgl,
         t.ljqfgyhs,t.ljfgyhs,t.xslbkstqs1,
         t.xslbkstqs2,t.xslksyhs,t.xslzcyhs_x,
         t.xslksyhs_x,
         IF (t.rq = ${v_date}, g.qryhs, t.qryhs)
         AS qryhs, t.gxdyyhs
    FROM tj_tqxsqk_r t LEFT OUTER JOIN (
         SELECT SUM(k.tqyhs) AS qryhs,
                k.tjrq,k.glfs, k.zjfs, k.dwdm
           FROM tj_tqxs_r k
          WHERE k.sfqr = 1
          GROUP BY k.tjrq,k.glfs, k.zjfs,k.dwdm) g
         ON t.rq = g.tjrq AND g.glfs = t.glfs
         AND g.zjfs = t.cjfs AND g.dwdm = t.dwdm
\end{lstlisting}

%

Unfortunately, in real enterprise data processing systems, data columns and
records are fairly big, while the number of columns and records that need to
be modified in one statement are limited. In our analysis, we found that most of the tables
in the smart grid system contain more than 50 columns, but the columns being
modified in one statement are less than 3 in average.
In most cases, the ratio of records that need to be modified is less than
1\%. Therefore, the INSERT OVERWRITE strategy will cause a large percentage of
redundant writes. For example, in an energy consumption table or partition
containing 1.8 billions records, only a few hundred records need to be modified for
a recurrent processing task.
Overwriting the whole table file will heavily degrade the efficiency of the
query.


Second, upgrading devices or modifying user information leads to change of archive
data as shown in Figure \ref{SystemArch} (2). In the Zhejiang Grid information system,
even in extreme cases, there are no more than 500 out of 22 million devices
changed on a single day. 
Thus the ratio of upgrading devices information is also very small. However, it
takes more than 15 minutes to rewrite the device information in the  Cloud Data Storage
System if Hive's overwrite operation is utilized.

Finally, the update operations of data recollection shown in Figure \ref{SystemArch} (1) also only affect a very small
amount of data, approximately less than 2000 records in a single update operation,
which yields an update ratio of less than 0.01\%. However, rewriting the electricity
consumption table in the Cloud Data Storage System will take more than half an hour
with our current cluster setting.


Apache Hive also plans to support data modification operations \cite{Hive-5317}. 
We will give a detailed comparison to our technique in Section \ref{sec:impl}.
All these instances reflect the same general problem of the lack of efficient update
operations in Hive. In the next section, we present DualTable, our solution to this problem.

\section{DualTable} \label{sec:dualtable}

\begin{figure}
 \centering
 \includegraphics[width=0.9\columnwidth]{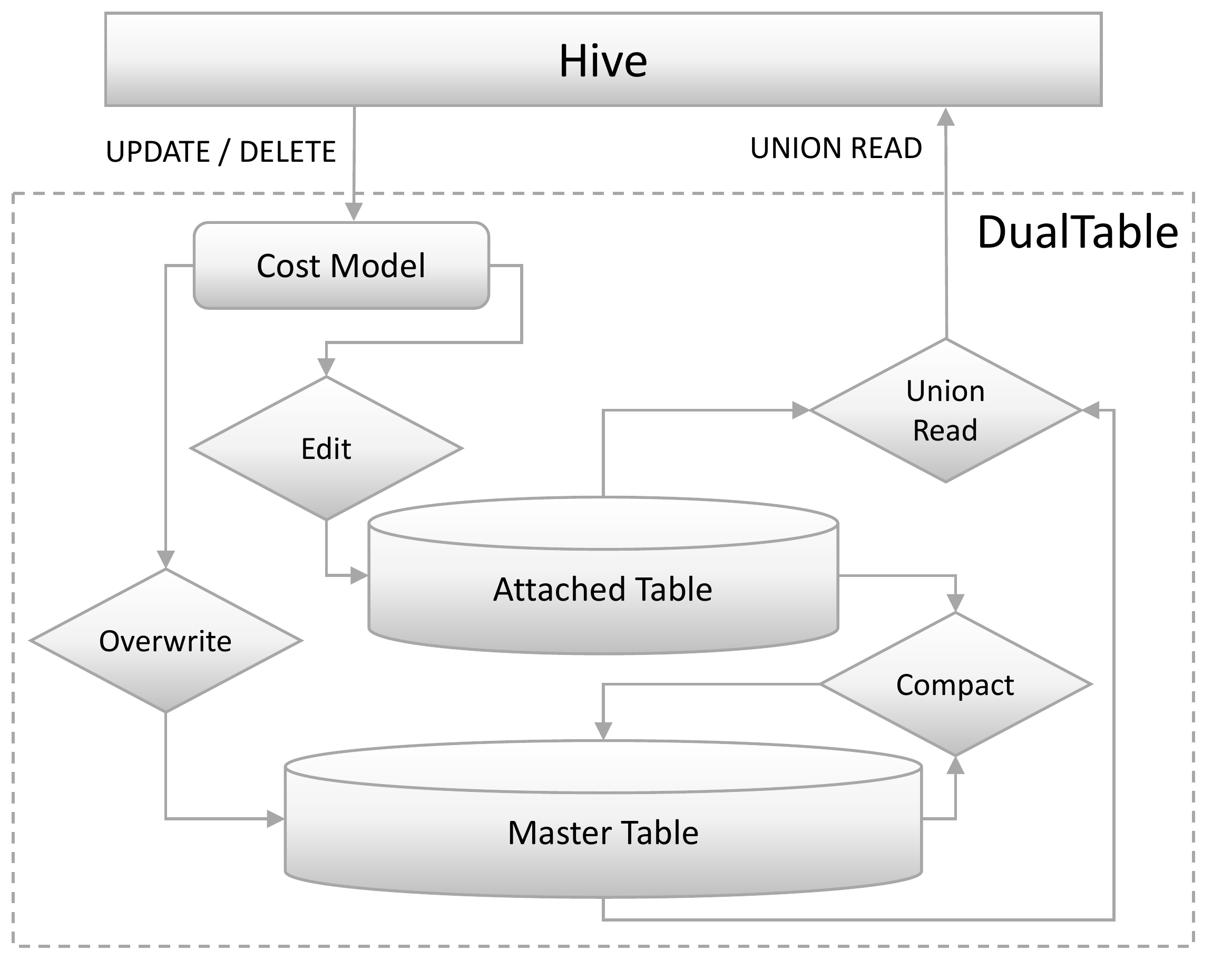}
 \caption{DualTable Architecture}
 \label{fig:dualtable}
\end{figure}

%
%
DualTable is a system that combines the strong read performance of HDFS with the update
performance of HBase. An abstract view of the architecture can be seen in Figure
\ref{fig:dualtable}. Data is stored in two locations, the \emph{Master Table} and the \emph{Attached Table}.
The Master Table is the main data storage, it is optimized for batch reading
and initially contains all the records in the table;
the Attached Table is an additional storage location for maintaining information about the updated or
deleted records.
 Besides Hive's original data manipulation operations INSERT INTO, CREATE, DROP,
and LOAD, DualTable also supports the additional operations UNION READ, UPDATE, DELETE and COMPACT;
 the \emph{Cost Model} is used to select an implementation plan for
UPDATE and DELETE operations and it includes separate cost models for each operation.

Each DualTable contains one Master Table and one Attached Table. Each row
recorded in DualTable has a unique ID in the table scope, which links the record
data in the Master Table and the Attached Table. When an UPDATE or DELETE
operation is executed, the system will choose an implementation plan based on the
cost model, either an \emph{OVERWRITE Plan} or an \emph{EDIT Plan}. The
OVERWRITE Plan rewrites the Master Table using Hive's original INSERT OVERWRITE
syntax while the EDIT Plan writes modification information to the Attached Table.
In order to combine the data from the Master Table and the Attached table, a
UNION READ operation is used, it generates a merged view. When the Attached Table
is too large and the merging becomes too expensive, the Attached Table is
compacted into the Master Table and cleared using the COMPACT operation.

\subsection{Master Table}

The Master Table stores the main part of the data. The storage used for this table
must provide high performance batch read and write. It can be implemented using HDFS,
Google File System \cite{Ghemawat2003}, the Quantcast File System \cite{QFS}, or optimized file formats
such as ORC \cite{Leverenz2013}.


\subsection{Attached Table}

The Attached Table stores information about the new values for updated record
fields or delete markers for deleted records. All these record modification
data are associated with their record IDs so that they can be merged with the
record data in the Master Table. The storage used for this table should provide
high performance random read and write. Possible candidates are, for example,
HBase, MySQL, and MongoDB. Due to the good integration, we will only discuss the
HBase-based implementation. 

\subsection{DualTable Operations}

Below we will characterize the basic storage operations in DualTable.

\begin{itemize}
  \item \textbf{CREATE and DROP:} Using the CREATE operation, DualTable
  will create an Attached Table and a Master Table. Analogously, it
  will delete the Master Table and the Attached Table when a DROP operation
  is issued.
  \item \textbf{LOAD and INSERT:} LOAD and INSERT are the same operations
  as in original Hive. Data are loaded and inserted into the Master Table. A
  unique ID will be assigned to each Master Table file, which is necessary
  to generate a unique ID for each row record. See Section \ref{sec:impl}
  for further details.
  \item \textbf{UPDATE and DELETE:}  When an UPDATE or DELETE operation is
  issued, the cost model will be used to choose the most efficient implementation
  plan from OVERWRITE Plan and EDIT Plan. The OVERWRITE Plan will execute
  Hive's INSERT OVERWRITE and replace the existing Master Table and Attached
  Table with a newly generated Master Table and an empty Attached Table, while
  the Edit Plan will add the updated information into the Attached
  Table. For an UPDATE operation, it will add the new value for the updated record
  fields. For a DELETE operation, it will add a DELETE marker to a corresponding
  record ID. In both UPDATE and DELETE the Master Table will not be changed.
  \item \textbf{UNION READ:} UNION READ reads and merges data from Master Table
  and Attached Table using the record ID. In order to make the merge process
  more efficient, we keep record IDs in the Master Table and Attached Table sorted,
  and implement a simple Map Reduce algorithm using a divide-and-conquer strategy.
  \item \textbf{COMPACT} As more data modification information is stored in the
  Attached Table with the EDIT Plan, the Attached Table grows. The more
  data is in the Attached Table, the higher the cost of the UNION READ operation,
  since it needs to read and merge data in the Master Table and the Attached
  Table. COMPACT does a UNION READ through the existing tables and creates a new
  Master Table by using INSERT OVERWRITE operation, which replaces the existing Master Table and Attached Table.
  All the other operations will be blocked during COMPACT.
  COMPACT can be scheduled to off-line hours or issued manually if the cost of
  a UNION READ is too expensive.
\end{itemize}

\section{Cost Model} \label{sec:costmodel}
DualTable uses a cost model to choose the most efficient implementation plan, OVERWRITE or EDIT,
for UPDATE and DELETE operations.
The cost of a plan consists of two parts:
\begin{enumerate}
  \item cost of reading and writing the Master Table
  \item cost of reading and writing the Attached Table.
\end{enumerate}

To determine the best implementation, the cost model estimates the costs of
both OVERWRITE and EDIT by computing the cost of data reading and writing separately.
By subtracting one from the other, the best plan can be found. If the result is positive,
it means that EDIT plan is cheaper and thus it will be chosen. Otherwise, the OVERWRITE
plan will be used. To calculate the costs, we make
the following assumptions: 

\paragraph{Notation 1} In a storage table $S$, the cost of reading or
writing data of the amount $D$ is denoted as $\mathcal{C}_\text{Read}^{S}(D)$
and $\mathcal{C}_\text{Write}^{S}(D)$, where $S$ can be $M$ (Master Table) or
$A$ (Attached Table).

\paragraph{Assumption 1} We assume that the cost of reading and writing is directly
proportional to the data volume read/written. This is denoted as
$\mathcal{C}_\text{Read}^{S}(\lambda D) \approx \lambda\mathcal{C}_\text{Read}^{S}(D)$, where $\lambda \in (0,1)$. The same holds for $\mathcal{C}_\text{Write}^{S}(D)$.

\paragraph{Notation 2} The total cost of a plan $P$ is denoted as $\text{Cost}_P$,
where $P$ can be OVERWRITE or EDIT.

\paragraph{Assumption 2} $\text{Cost}_P$ equals "modification cost" plus
"following read cost", where "modification cost" indicates the total cost
to execute UPDATE or DELETE using plan $P$; "following read cost" indicates
the cost to read the whole table for $k$ times after UPDATE or DELETE completes.

Given a DualTable $T$ containing data of size $D$, suppose we execute one
modification on $T$ and then read the table $k$ times, the
corresponding cost models for UPDATE and DELETE are illustrated as follows.

\paragraph{UPDATE Cost Model}
Suppose the ratio of data updates is denoted as $\alpha$, $\alpha \in (0,1)$,
the costs of the OVERWRITE plan and EDIT plan are shown below, each consisting of two parts:

\begin{center}
\resizebox{\columnwidth}{!}{%
\begin{tabular}{lcc}
 & Update Cost & Following Read Cost \\ \hline
OVERWRITE & $\mathcal{C}_\text{Write}^{M}(D)$ & $k\mathcal{C}_\text{Read}^{M}(D)$ \\ \hline
EDIT      & $\mathcal{C}_\text{Write}^{A}(\alpha D)$ &  $k(\mathcal{C}_\text{Read}^{A}(\alpha D) + \mathcal{C}_\text{Read}^{M}(D))$
\end{tabular}}
\end{center}

Let $\text{Cost}_U$ be the cost of OVERWRITE plan minus the cost of EDIT plan:

\begin{eqnarray}
\text{Cost}_U &=& \text{Cost}_\text{OVERWRITE} - \text{Cost}_\text{EDIT}    \nonumber \\
              &=& \mathcal{C}_\text{Write}^{M}(D) + k\mathcal{C}_\text{Read}^{M}(D) - \mathcal{C}_\text{Write}^{A}(\alpha D)  \nonumber \\
              &-& k(\mathcal{C}_\text{Read}^{A}(\alpha D) +\mathcal{C}_\text{Read}^{M}(D))  \nonumber \\
						  &=& \mathcal{C}_\text{Write}^{M}(D) + k\mathcal{C}_\text{Read}^{M}(D) - \alpha\mathcal{C}_\text{Write}^{A}(D)  \nonumber \\
              &-& k\alpha\mathcal{C}_\text{Read}^{A}(D) - k\mathcal{C}_\text{Read}^{M}(D) \nonumber \\
							&=& \mathcal{C}_\text{Write}^{M}(D) -  \alpha(\mathcal{C}_\text{Write}^{A}(D) + k\mathcal{C}_\text{Read}^{A}(D))
\label{eq:costu}
\end{eqnarray}

The update ratio $\alpha$ can be estimated using historical analysis of the
execution log or can directly be given by the designer. The number of successive read operations
after an update $k$ can directly be set by the designer, or inferred from the
HiveQL code.


Using the model, it is clear that when $\alpha$ and $k$ is small, $\text{Cost}_U$
can be positive. This means that the EDIT plan is more efficient when the update ratio
and the number of consecutive reads are small. On the other hand, when the update
ratio and the number of consecutive reads become too large, the OVERWRITE plan
is a better choice.

As an example, suppose we use HDFS for hosting the Master Table $M$ and HBase
for the Attached Table $A$, the data volume $D$ = 100GB, update ratio
$\alpha=0.01$. The rate of HDFS writes using multiple Map tasks adds up to 1GB/s.
The rate of HBase reading and writing add up to 0.5GB/s and 0.8GB/s, respectively.
Suppose we read continuously for 30 times after the updating operation, the cost
model can be computed as follows:

\begin{eqnarray}
\text{Cost}_U &=& \text{Cost}_\text{OVERWRITE} - \text{Cost}_\text{EDIT}    \nonumber \\
							&=& \mathcal{C}_\text{Write}^{M}(D) -  \alpha(\mathcal{C}_\text{Write}^{A}(D) + k\mathcal{C}_\text{Read}^{A}(D))   \nonumber \\
							&=& 100GB / 1GBps - 0.01 \cdot (100GB/ 0.8 GBps \nonumber \\
                                   &+& 30 \cdot 100GB/0.5GBps) \nonumber \\
							   &=& 38.75s \nonumber
\label{eq:costuexample}
\end{eqnarray}

As in this example, the time consumption of EDIT plan is shorter than that of OVERWRITE plan. We will choose EDIT as a consequence.

\paragraph{DELETE Cost Model}

Suppose the ratio of records being deleted is $\beta$ and $\beta \in (0,1)$.
Suppose the average data size of each row is $d$, the size of a DELETE marker is $m$,
then the data size of deleted data, denoted as $\beta D$, is $\frac{\beta D}{d}m$.
The cost of OVERWRITE and EDIT plans are shown below:
\begin{center}
\resizebox{\columnwidth}{!}{%
\begin{tabular}{lcc}
 & DELETE Cost & Following Read Cost \\ \hline
OVERWRITE & $\mathcal{C}_\text{Write}^{M}((1-\beta)D)$ & $k\mathcal{C}_\text{Read}^{M}((1-\beta)D)$ \\ \hline
EDIT     & $\mathcal{C}_\text{Write}^{A}(\frac{\beta Dm}{d})$ &  $k(\mathcal{C}_\text{Read}^{A}(\frac{\beta Dm}{d}) + \mathcal{C}_\text{Read}^{M}(D))$
\end{tabular}}
\end{center}

Let $\text{Cost}_D$ be cost of OVERWRIT plan minus the cost of EDIT plan.
It can be computed as follows:
\begin{eqnarray}
\text{Cost}_D &=& \text{Cost}_\text{OVERWRITE} - \text{Cost}_\text{EDIT}    \nonumber \\
              &=& \mathcal{C}_\text{Write}^{M}((1-\beta)D) + k\mathcal{C}_\text{Read}^{M}((1-\beta)D)     \nonumber \\
              &-& \mathcal{C}_\text{Write}^{A}(\frac{\beta Dm}{d}) - k(\mathcal{C}_\text{Read}^{A}(\frac{\beta Dm}{d}) + \mathcal{C}_\text{Read}^{M}(D))  \nonumber \\
						  &=& (1-\beta)\mathcal{C}_\text{Write}^{M}(D) + \beta k\mathcal{C}_\text{Read}^{M}(D) - \beta\mathcal{C}_\text{Write}^{A}(D) \nonumber \\
              &-& k\beta\mathcal{C}_\text{Read}^{A}(\frac{Dm}{d}) - k\mathcal{C}_\text{Read}^{M}(D)  \nonumber \\
							&=& \mathcal{C}_\text{Write}^{M}(D) -  \beta(\mathcal{C}_\text{Write}^{M}(D) + k\mathcal{C}_\text{Read}^{M}(D)     \nonumber \\
              &+& \frac{m}{d}\mathcal{C}_\text{Write}^{A}(D)+k\frac{m}{d}\mathcal{C}_\text{Read}^{A}(D))
\label{eq:costd}
\end{eqnarray}

Where $m$ is a constant value, which can be determined via data sampling. Estimation
of $\beta$ is similar to that of $\alpha$ in the UPDATE cost model.
$\mathcal{C}_\text{Write}^{M}(D)$, $\mathcal{C}_\text{Write}^{A}(D)$, and
$\mathcal{C}_\text{Read}^{A}(D)$ can be computed the same way as in UPDATE cost model.

Using the model, it is obvious that when $\beta$ and $k$ are small, $\text{Cost}_D$
is positive. This means that the EDIT plan is more efficient when the delete ratio and
the consecutive number of reads are small. On the other hand, when the delete ratio
and the consecutive number of reads become larger, the OVERWRITE plan becomes more efficient.

\section{Implementation Details} \label{sec:impl}

\begin{figure}
 \centering
 \includegraphics[width=0.8\columnwidth]{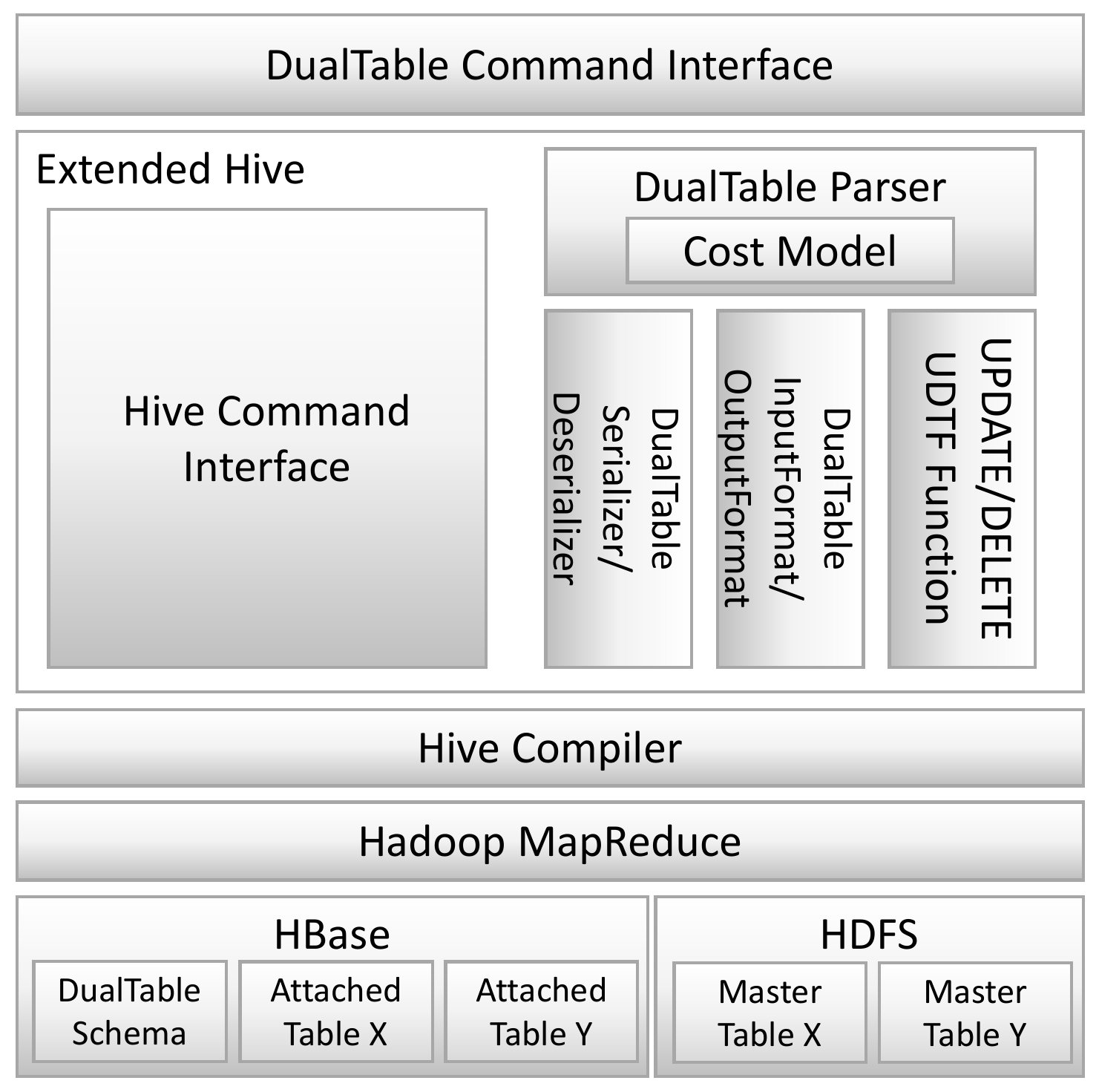}
 \caption{DualTable Implementation on Hive}
 \label{fig:dtimpl}
\end{figure}

We have implemented DualTable with Apache HBase, HDFS and Hive.
In this section, we discuss technical details about our extensions to Hive, the
data layout and the record ID management.

\subsection{Extensions to Hive}
Hive provides multiple abstractions that enable extensions: the InputFormat and OutputFormat
classes are used in a MapReduce job to read and write data rows. Hive uses
Serializer and Deserializer classes to parse records from data rows. Hive
supports \emph{user defined table functions} (UDTF) to add new functionality
in statements to manipulate data.

As shown in Figure \ref{fig:dtimpl}, we use HDFS for the Master Tables and
HBase for the Attached Tables and a system wide metadata table. Each DualTable
contains an HDFS-based \emph{Master Table} and an HBase-based \emph{Attach Table}.

We developed custom InputFormat, OutputFormat, Serializer, and Deserializer classes with
UNION READ and record ID management logic for DualTable. Additionally, two UDTFs implement
the EDIT Plans for UPDATE and DELETE. The UPDATE UDTF takes the name of the updated table,
the updated columns and the new values as input and stores the update information in HBase.
The DELETE UDTF only takes the name of the table and puts a DELETE marker for each deleted row in HBase.


We have added UPDATE and DELETE commands to HiveQL. If a HiveQL statement contains
an UPDATE or DELETE command, it will be sent to the DualTable parser, otherwise,
it will go to the original Hive parsing procedure.
For these two DML commands the parser will choose to generate a Hive-compatible statement using
INSERT OVERWRITE or our UDTFs, based on the cost evaluator. The former one is for
an OVERWRITE plan, and the later one for an EDIT plan.

The cost evaluator is in charge of cost evaluation based on our cost model
as described above. The DualTable metadata manager collects and manages information
required for cost evaluation.


\subsection{Data Layout and Record ID Management}
We use the ORC file format in HDFS for the Master Table and one Master Table
may consist of multiple ORC files in an HDFS folder. Besides the ORC file
format's handy features like compression and Hive type support, we chose it
for two important reasons:
\begin{enumerate}
 \item 
 We maintain an incremental integer
 file ID for each DualTable in the system wide metadata table. Whenever a MapReduce
 mapper creates a new 
 file, it retrieves and stores a unique ID in the 
 file
 metadata.
 \item We can retrieve row numbers when reading data rows.
 The row
 numbers are computed during reading operations and have no storage cost, which makes
 it a perfect way to maintain unique IDs for each DualTable row record. A DualTable
 record ID is generated on read by concatenating the 
 file ID and record's 
 row number, which makes the record ID unique in one DualTable.
\end{enumerate}
In the HBase-backed Attached Table, we use DualTable record IDs as HBase row keys. For
UPDATE information, the updated field's column number (as maintained by Hive) serves
as HBase column qualifier and the new field value as HBase cell value. For DELETE
information, only a delete marker (a special HBase cell) is stored in the deleted
record's ID row.

With the data layout and record ID generation policy above, sequential record IDs within
an ORC file are in ascending order. Meanwhile, record IDs stored as row keys in HBase are
already sorted. This makes it simple and straightforward for a Mapper to merge data in
the Master Table and the Attached Table for UNION READ operations because it only needs
to read through and merge two sorted ID lists.

\subsection{Comparison to Hive ACID Extensions}\label{sec:HiveACIDvsDualTable}
Apache Hive also plans to support data modification operations. They published a design document in 2013, but the up-to-date version Hive-0.13, which is released in April 2014, does not support data update or delete yet. The feature is still under development \cite{Hortonworks_trans}. Due to the fact that Hive-0.13 does not support UPDATE/DELETE statement, we could only compare the two systems from conceptual perspective.

DualTable puts the data modification information into a HBase table, which is called Attached Table; The original data is saved into Master Table; Each Master Table has only one Attached Table. The read operation accesses both Master Table and Attached Table to get the original data and its modification information, then combines them to get the up-to-date data view; For write operation (UPDATE or DELETE), DualTable could either overwrite the whole Master Table or just update the Attached Table, and it makes use of a cost-model to make decision. When the size of Attached Table exceeds a threshold, DualTable merges it with its Master Table.

Hive puts both the original data and modification information into the same Hive database \cite{Leverenz2014}. They are called base table and delta tables, respectively. Each transaction creates a new delta table for a base table. Therefore, a base table could have multiple delta tables. The read operation retrieves a record from base table and merge sorts it with corresponding records in delta tables to get the up-to-date data view. The write operation puts the whole updated record into delta tables, even if only one cell is changed. Hive supports two compact modes, minor compact merges all delta tables belonging to the same base table into a single delta table; and major compact merges the delta tables with their corresponding base table.

We compare DualTable and Hive from three aspects:
First, their objectives are different. Hive aims to support transaction and full ACID guarantee \cite{Leverenz2014}\cite{Hortonworks_ACID}. DualTable focuses on optimization of data update performance for our smart grid industrial scenarios.

Secondly, their storage policies are different. DualTable employs hybrid storage architecture to make full advantage of both HDFS and HBase. In this way, DualTable could improve random write performance significantly without obvious negative impact on sequential read. While Hive puts both the original data and modified information into HDFS. For data read operation, Hive merge sorts the base table with all relevant delta tables to get the up-to-date view. Since delta table is stored as plain Hive tables and updated records are all appended to the tables, the reader has to scan them sequentially and selects latest updated values for particular record. On the contrary, DualTable retrieves a row from master table, then randomly accesses HBase based Attached table to get changed record and its latest value according to the row ID. They are combined in the UnionRead operation. In addition, DualTable can make use of HBase's multiple-version feature to track data change history.

Third, DualTable supports runtime selection of update policy. Our experiments find that overwriting the whole table with INSERT OVERWRITE statement sometimes performs better when update ratio exceeds a threshold. Therefore, DualTable incorporates a cost model to decide whether to put data modification information into the Attached Table or overwrite the whole table. However, Hive always updates the delta tables. It could not make better decisions at runtime.

\section{Evaluation} \label{sec:eval}
In following sections, we compare DualTable with Hive in terms of query performance
and performance of update and delete operations by experiments. 



We conduct two sets of experiments. The first set of experiments uses a dataset from the Zhejiang Grid
and runs on a cluster of 26 nodes; In order to further assess the generic applicability of DualTable, we perform the second set of experiments with TPC-H dataset on a 10-node cluster.
Each node is equipped with 8 cores, 16 GB memory, and 250 GB hard disk. All nodes run CentOS 6.2, Java
1.6.0-41, Hadoop-1.2.1 and HBase-0.94.13. We implement DualTable based on
Hive-0.11. Since DualTable is implemented based on ORC file format, we set Hive to use the same file format for
fair comparison. JobTracker, Namenode and HMaster run on the same node. TaskTracker,
Datanode, and RegionServer run on other nodes. Every worker in Hadoop is
configured with up to 6 mappers and 2 reducers. HDFS is configured with 3 replicas
and 64 MB chunk size. We run all experiments three times and report the average
result. The metric used in our experiment is run time.

\subsection{Evaluation of Real Grid Workloads}
We carried out our performance evaluation with production data collected from
electricity information collection system deployed in the Zhejiang Grid 
of the China State Grid, which is the largest electric utilities company in the country.
In order to make the experiment easy to conduct and the run time controllable,
the total data set we use is around 64 GB.
To avoid use of memory cache, we reset the system every time when we finish one experiment.

Since the IO cost of Hive will increase
nearly linearly with the growth of data size, it is obvious that the performance
trend using this workload is typical and will reflect the trend in bigger or smaller
workloads. The six tables involved are listed in Table \ref{tab:GridSchema}. We also list some representative
columns involved in our experiments.
We have also tested HBase-based Hive, which can also
support update and delete operations rather than Hive's default INSERT OVERWRITE
operation. The TPC-H workload running on a 10-node setting shows that HBase-based
Hive is much slower than Hive itself and DualTable, respectively. This is the reason
why we do not consider HBase-based Hive as a comparison target system in this section.

\begin{table}
\begin{center}
\begin{tabularx}{\columnwidth}{l|r|X}
Table & \# Records & Columns in Experiments \\ \hline
yh\_gbjld	& 7112576	& dwdm: organization code; gddy: voltage; hh: family id; sfyzx: withdrawn or not \\
zd\_gbcld	& 7963648 & cldjh: measure point id; zdjh: terminal code; dwdm: organization code; \\
zc\_zdzc	& 74104736 & dwdm: organization code; zdjh: terminal code; zzcjbm: manufacture code; cjfs: collection method; zdlx: terminal type; \\
rw\_gbrw	& 34045664 & xfsj: issued time; rwsx: task property; cldh: measure point id; \\
tj\_gbsjwzl\_mx &	239032928	& yhlx: user type; rq: date; dwdm: organization code; cjbm: manufacture code; \\
tj\_dzdyh & 9805312 & zdjh: terminal code;
\end{tabularx}
	\caption{Schema Excerpt of the Real State Grid Data Set}
\label{tab:GridSchema}
\end{center}
\end{table}

\begin{figure*}[t!]
\begin{center}
\begin{minipage}[b]{0.3\textwidth}
 \centering
 \includegraphics[width=\columnwidth]{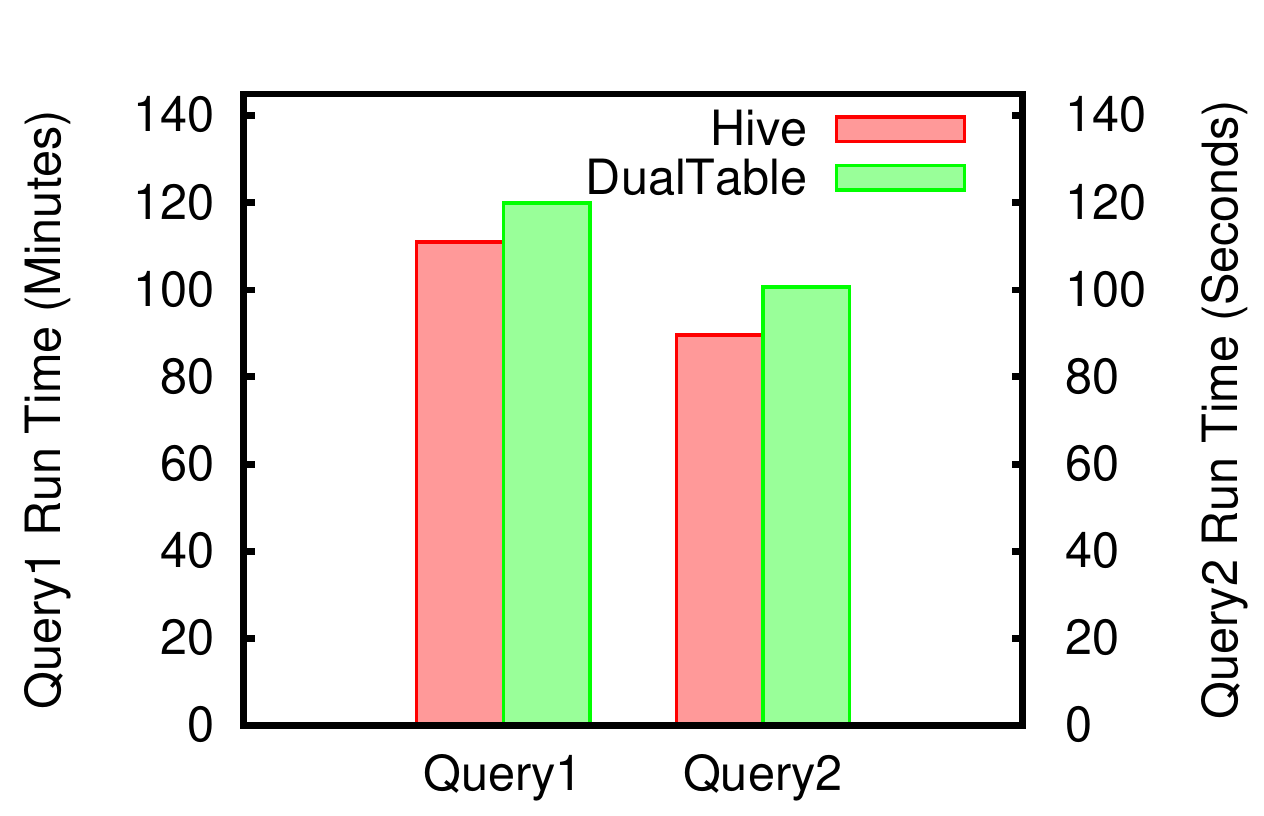}
 \caption{Read Performance Comparison of Hive and DualTable Select Statement 1 \& 2}
 \label{fig:Select1}
\end{minipage}%
\hspace{0.05cm}
\begin{minipage}[b]{0.3\textwidth}
 \centering
 \includegraphics[width=\columnwidth]{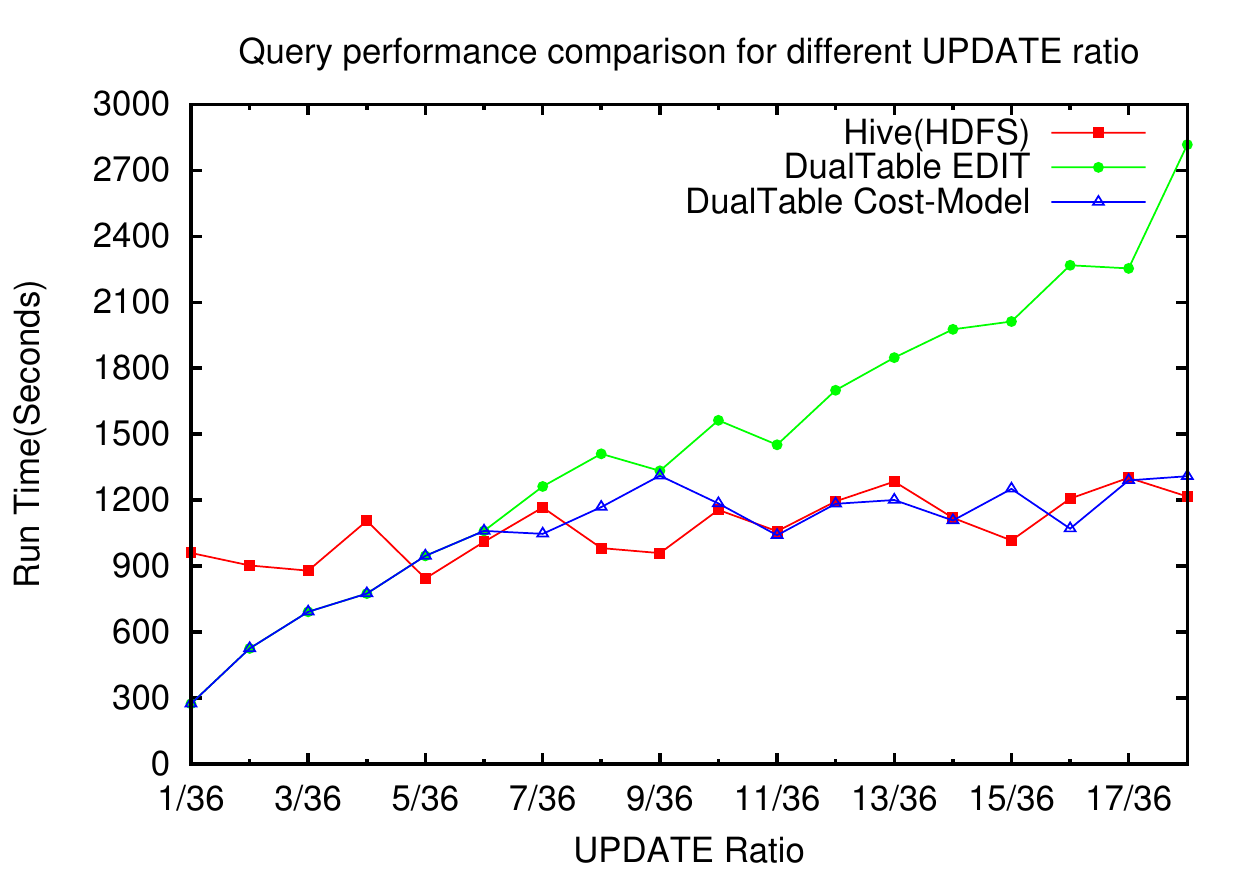}
 \caption{Update Performance Comparison for Various Data Modification Ratios}
 \label{fig:Update1}
\end{minipage}
\hspace{0.05cm}
\begin{minipage}[b]{0.3\textwidth}
 \centering
 \includegraphics[width=\columnwidth]{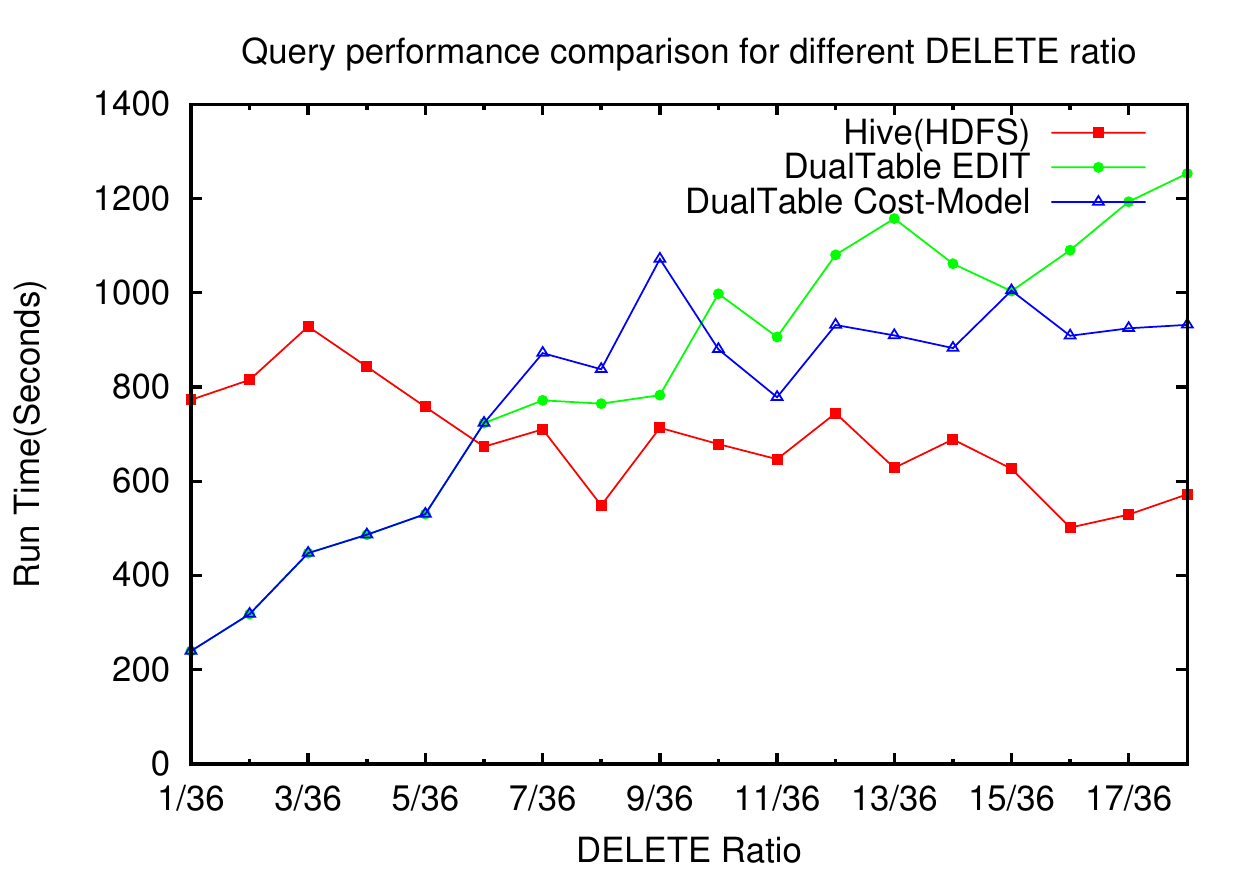}
 \caption{Delete Performance Comparison for Various Data Modification Ratios}
 \label{fig:Delete1}
\end{minipage}%

\begin{minipage}[b]{0.3\textwidth}
 \centering
 \includegraphics[width=\columnwidth]{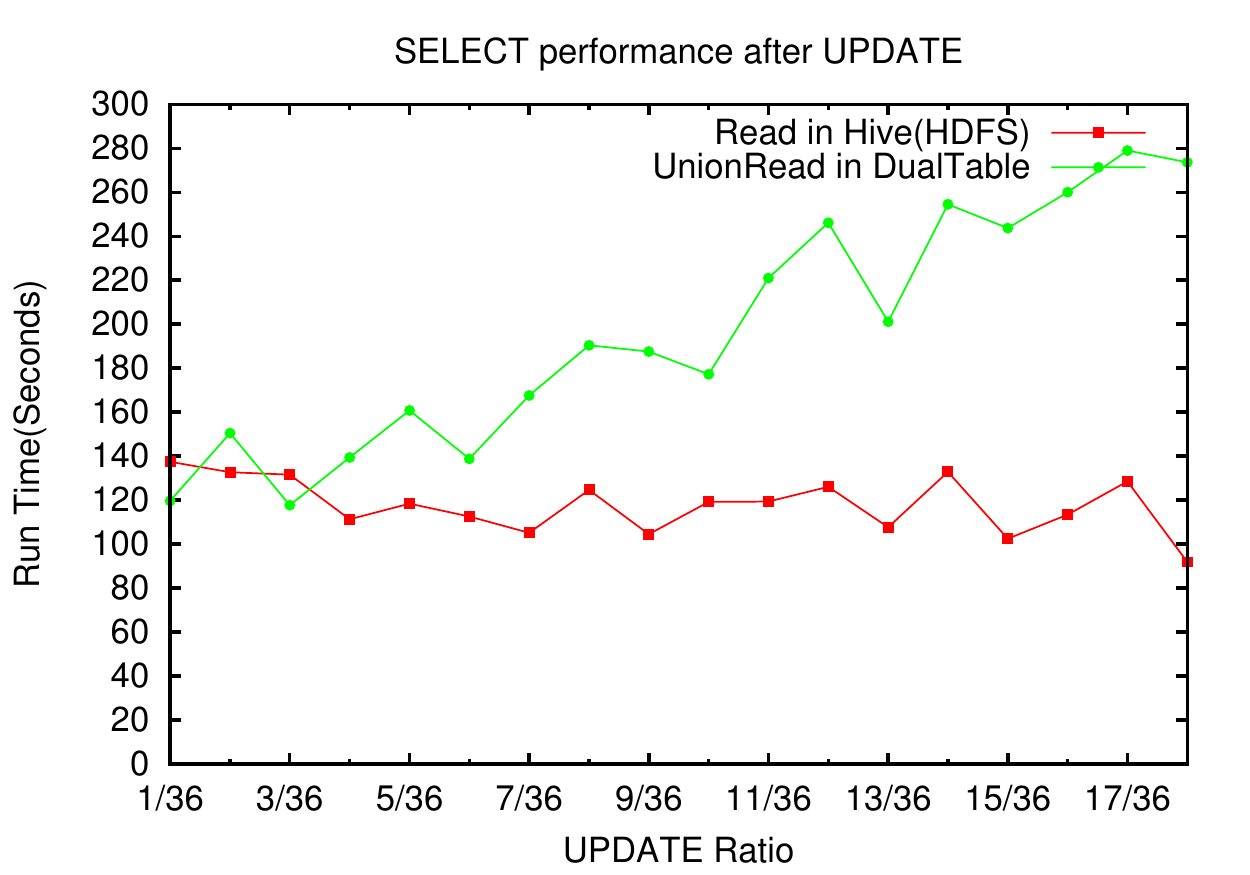}
 \caption{Run Time of a SELECT Query Following the UPDATE Operation}
 \label{fig:Update2}
\end{minipage}%
\hspace{0.05cm}
\begin{minipage}[b]{0.3\textwidth}
 \centering
 \includegraphics[width=\columnwidth]{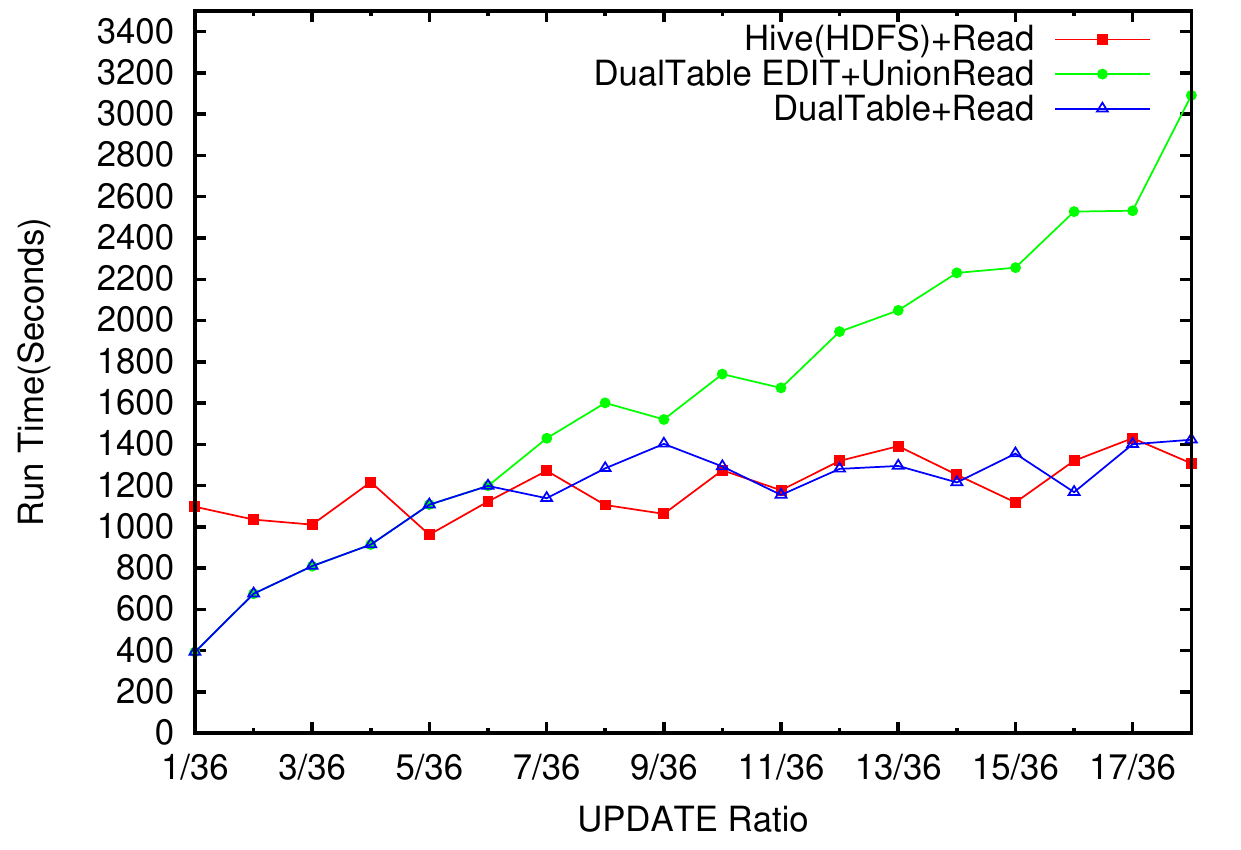}
 \caption{Total Run Time of a SELECT Query Following the UPDATE Operation}
 \label{fig:UpdateSelectTotalTime}
\end{minipage}
\hspace{0.05cm}
\begin{minipage}[b]{0.3\textwidth}
 \centering
 \includegraphics[width=\columnwidth]{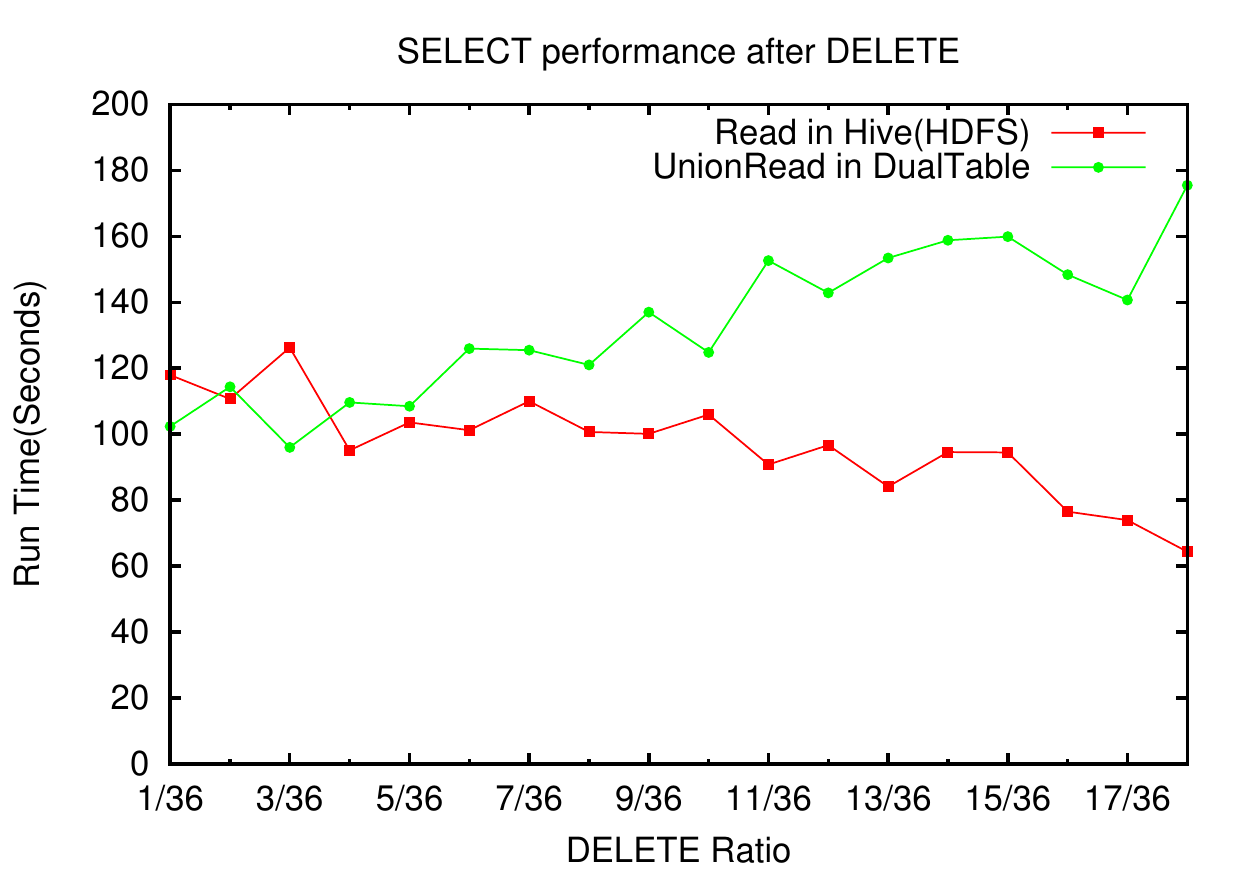}
 \caption{Run Time of a SELECT Query Following the DELETE Operation}
 \label{fig:Delete2}
\end{minipage}%
\end{center}
\end{figure*}

{\bf Performance Overhead of Queries:} In the first experiment, we assess read performance of DualTable and Hive using
two typical SELECT statements of State Grid business logic. The first statement
retrieves records from table yh\_gbjld according to some predicates, in which
yh\_gbjld joins with table zc\_zdzc and table zd\_gbcld. The second statement
calculates total number of records in table tj\_gbsjwzl\_mx. The Attached Table
of DualTable is empty in this experiment. Both Hive and DualTable scan the whole
table to filter records. Since the Attached table is empty, DualTable does not
need to merge the original record from Master Table with data modification information.
Figure \ref{fig:Select1} shows the results. For statement \#1, Hive takes 111 minutes and
DualTable takes 120 minutes. The performance difference is about 8\%, which is
attributed to the overhead incurred by the Attached Table (although it does not
contain any data, the function invocation is inevitable).  For Statement \#2,
Hive takes 89 seconds and DualTable takes 101 seconds.
Hive outperforms DualTable about 12\%, which again is attributed to overhead of
the Attached Table. This experiment shows that the overhead of the Attached Table is fairly
low.

%




{\bf Performance of Updates:} This experiment demonstrates how Hive and DualTable perform when handling update
and delete operation. It is very common in the business logic of State Grid to
change or remove records of some specific dates. We mimic this behavior in this
test. The tables involved contain roughly uniformly distributed data of 36 days,
and the experiment starts by changing data of one day ($\frac{1}{36}$) until data of
18 days ($\frac{18}{36}$). In order to verify the effectiveness of the cost model, we first
run DualTable with cost-model and, as a comparison, DualTable in EDIT mode, which
means DualTable always writes data modification information into the Attached Table.
Figure \ref{fig:Update1} shows the performance of DualTable and Hive for an update operation.
It can be seen that Hive's execution time does not fluctuate much with variation
of data modification ratio, since Hive always overwrites the whole table. For
DualTable, the cost of writing update information into the Attached Table is
proportional to the amount of updated data. When the update ratio is smaller
than $\frac{6}{36}$, the cost model selects \emph{EDIT} instead of \emph{OVERWRITING},
so \emph{DualTable EDIT} overlaps with \emph{DualTable cost-model}; the cost of
writing into Attached table is less than overwriting the whole data, which makes
DualTable perform significantly better than Hive. When data update ratio
increases, the execution time of DualTable EDIT mode grows drastically. When
data update ratio exceeds $\frac{6}{36}$, DualTable switches to \emph{OVERWRITE} mode,
and DualTable takes a little longer than Hive to run the UPDATE statement due
to its own overhead.
%

Figure \ref{fig:Delete1} depicts a performance comparison of delete operations
on Hive and DualTable with various data deletion ratios. Hive's \emph{overwrite
the whole table} approach results in reduction of data written into HDFS when
data deletion ratio increases, so its run time is inversely proportional to the
delete ratio. Hive's run time drops from 772 seconds to 572 seconds when the
ratio raises from $\frac{1}{36}$ to $\frac{18}{36}$. In the other hand, \emph{DualTable EDIT} puts
a \emph{DELETE marker} for each removed row into the Attached Table. As a result,
its run time increases with the data deletion ratio. When $\frac{1}{36}$ of the data is
deleted, DualTable outperforms Hive by a factor of 3. With a delete
ratio smaller than $\frac{10}{36}$, the cost model selects \emph{EDIT} instead of
\emph{OVERWRITE}. Therefore, \emph{DualTable EDIT} overlaps with \emph{DualTable cost-model}.
After that, the cost of writing DELETE markers into HBase exceeds the overhead
of overwriting the whole table, and DualTable starts to adopt the overwriting
approach to accomplish data deletion. There is a small overhead to run the DELETE statement.



{\bf Impact of Size of Attached Table:} The previous experiment evaluates update performance of DualTable and Hive with
various data modification ratios. We analyzed the State Grid workload and found
that changed tables will be retrieved in subsequent operations to get the latest
values. To reflect this in the experiments, we issue a SELECT query after
UPDATE and DELETE operations like we did in last experiment, to show how the size of
Attached table impacts performance of following UnionRead operations.



Figure \ref{fig:Update2} shows the run time of a SELECT query following the UPDATE operation
used in the previous experiment. Hive performance does not fluctuate much with the
UPDATE ratio, since the UPDATE operation does not change data amount in the related
Hive tables. In this experiment, DualTable is always slower than Hive. The
performance difference is very small when only one specific day's data is updated;
however, DualTable takes more time for UnionRead operation with raising UPDATE
ratio, and it is 2.7 times slower than Hive when the UPDATE ratio grows to 18/36.
This is because DualTable EDIT mode puts all UPDATE information into the Attached
Table, and the following UnionRead operation needs to first read the original
record from the Master Table, then merge with the corresponding record in the
Attached Table to get the latest value.
Figure \ref{fig:UpdateSelectTotalTime} demonstrates the total time taken by the UPDATE operation and the
following SELECT query.	The trend shown in this figure and its explanation is
similar to Figure \ref{fig:Update1} and, therefore, we do not repeat it here for space
limitation.



Figure \ref{fig:Delete2} and Figure \ref{fig:DeleteSelectTotalTime} depict the run time of a SELECT query following
the DELETE operation used in the previous experiment. These results are similar
to the last one we just explained, therefore, we do not repeat it here for space limitation.


\begin{table}
\begin{center}
\begin{tabularx}{\columnwidth}{l|r|X}
Table & \# Records & Columns in Experiments \\ \hline
tj\_tdjl & 58494976 & tdsj: outage time; qym: area code; zdjh: terminal code; \\
tj\_td & 33036288 & hfsj: recovery time; tdsj: outage time;\\
tj\_sjwzl\_r &73569360 & rq: date; rcjl: sampling rate of a day; yhlx: user type;\\
tj\_dysjwzl\_mx & 382890014 & rq: date; sfld: miss a point or not; cjfs: collection method;\\
tj\_sjwzl\_y & 2586120 & rq: date\\
tj\_gk & 30655920 & rq: date; dwdm: organization code;\\
\end{tabularx}
    \caption{Schema Excerpt of the State Grid Data Set}
\label{tab:GridSchema2}
\end{center}
\end{table}

\begin{table*}
\begin{center}
\begin{tabularx}{\textwidth}{l|X|r|r|r|r}
Stmt & Semantics & Update Ratio & Hive (sec) & DualTable (sec) & Improvement \\ \hline
U \#1 & Set the area code in which an outage event happens at some specified time to a new value.  &  2\% & 159.81 & 51.39 & 311\% \\
U \#2 & When the outage recovery time is earlier than the start time, set the outage recovery time to a value which indicates an error. & 5\% & 104.90 & 60.81 & 173\% \\
U \#3 & set the sampling rate of a day to a new value for a specified date and specified user type. & 0.1\% & 389.19 & 47.52 & 819\% \\
U \#4 & Set the collection method of a specified day and specified user type to a new value. & 3\% & 1577.87 & 161.73 & 976\% \\
D \#1 & Delete records from table tj\_sjwzl\_y for a specified month. & 4\% & 46.26 & 22.47 & 206\% \\
D \#2 & Delete records from table tj\_tdjl for a specified area code. & 5\% & 102.04 & 47.26 & 216\% \\
D \#3 & Delete records from table tj\_gk for a specified organization code and a marker. & 3\% & 147.87 & 34.97 & 423\% \\
D \#4 & Delete records from table tj\_tdjl for a specified terminal code and outage time. & 0.01\% & 140.94 & 29.47 & 478\%
\end{tabularx}
    \caption{Performance Results for Real State Grid Workload}
\label{tab:realResults}
\end{center}
\end{table*}

{\bf More Experiments:} As mentioned above, the data modification ratio is rarely higher than 10\% in the data
analysis system of the State Grid. In order to further verify the effectiveness of
DualTable regarding State Grid workload, we extracted four representative UPDATE
statements and DELETE statements from line loss and low voltage calculation modules.
The six tables involved are listed in Table \ref{tab:GridSchema2}. Their total size is 70 GB.
We also list some representative columns involved in the experiments. Their data
modification ratio ranges from 0.01\% to 5\%. We compare DualTable and Hive in
terms of query run time, and calculate the performance improvement of DualTable
in Table \ref{tab:realResults}(U is abbreviation of UPDATE, D is abbreviation of DELETE in the
table). We can see that DualTable outperforms Hive an order of magnitude for all
the 8 operations thanks to its cost model and the Attached Table storage model.

\subsection{TPC-H Workloads Evaluation}

Besides the above performance evaluation conducted with real production data,
we further assessed the generic applicability of DualTable using the
standard TPC-H queries and data.

We conduct a number of experiments to measure the read and update performance of DualTable.
When we use update or delete in HBase-based Hive, we implement the EDIT plan
similar to DualTable using user defined functions instead of relying on the INSERT OVERWRITE statement.

The tables lineitem and order of TPC-H were used, these are the two largest tables of the TPC-H data set. In the TPC-H 30GB
data set that was used, they have 0.18 billion rows (i.e., 23GB) and 45 million rows (i.e., 5GB) respectively.
We modify TPC-H queries to add update and delete operations.

\begin{figure*}[t!]
\begin{center}
\begin{minipage}[b]{0.3\textwidth}
 \centering
 \includegraphics[width=\columnwidth]{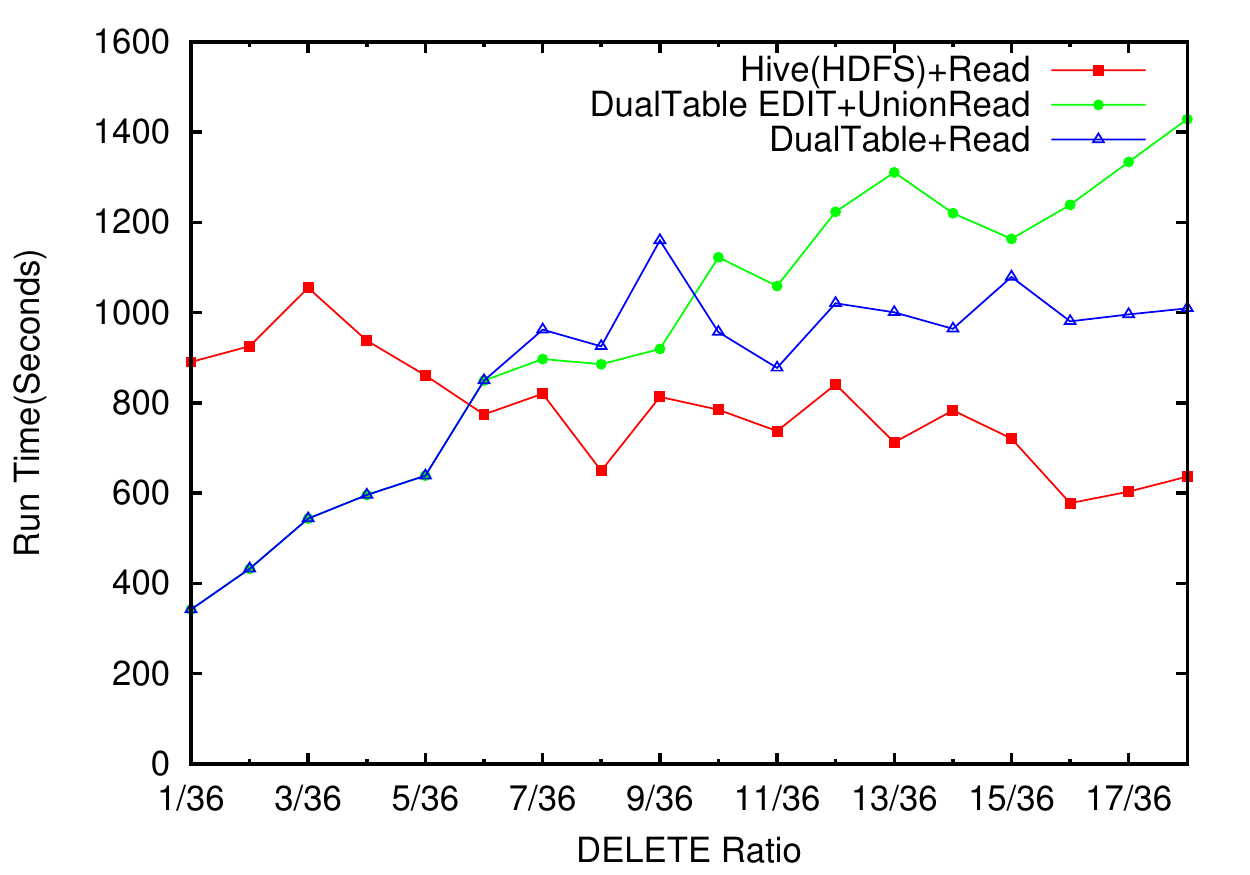}
 \caption{Total Run Time of a SELECT Query Following the DELETE Operation}
 \label{fig:DeleteSelectTotalTime}
\end{minipage}%
\hspace{0.05cm}
\begin{minipage}[b]{0.3\textwidth}
 \centering
 \includegraphics[width=\columnwidth]{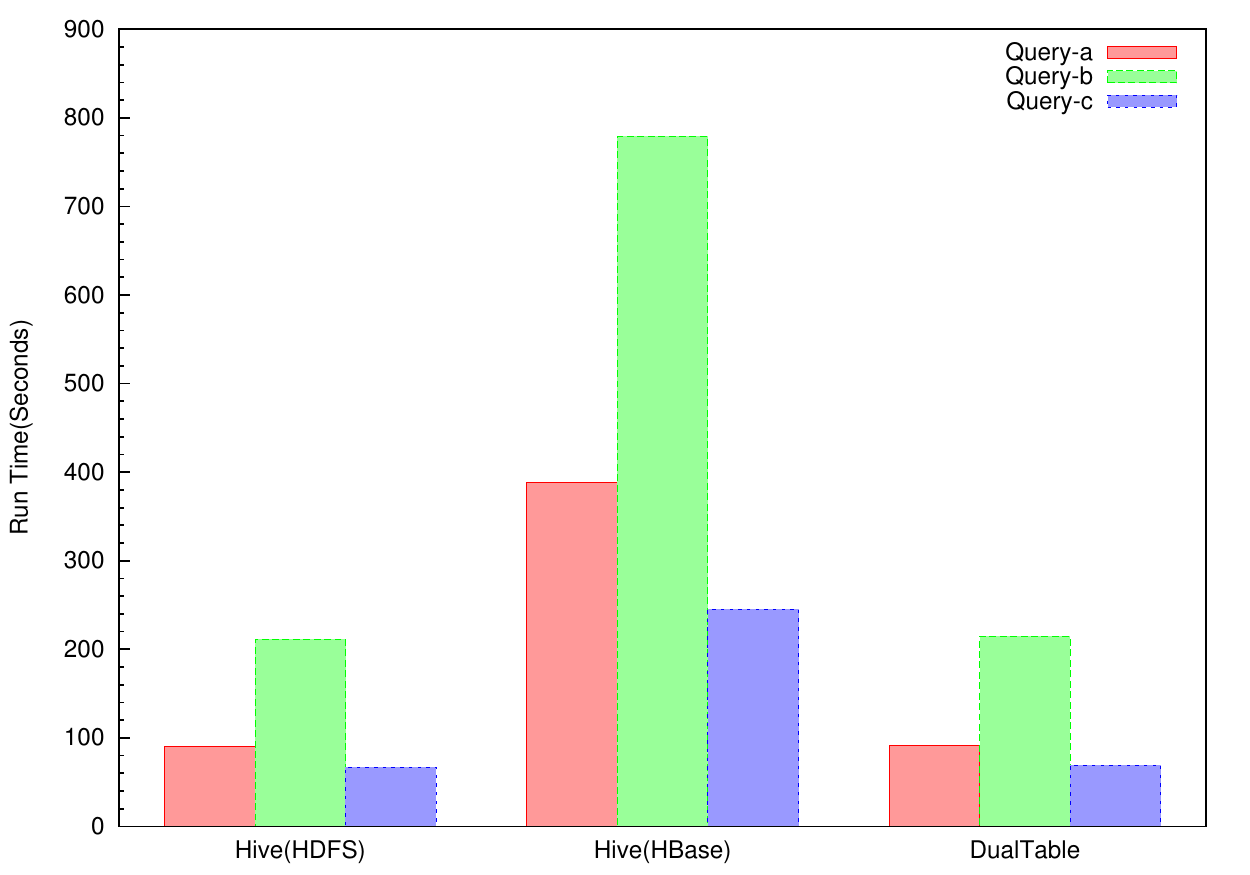}
 \caption{Read Performance on 30GB TPC-H Data Set}
 \label{fig:Chart1}
\end{minipage}
\hspace{0.05cm}
\begin{minipage}[b]{0.3\textwidth}
 \centering
 \includegraphics[width=\columnwidth]{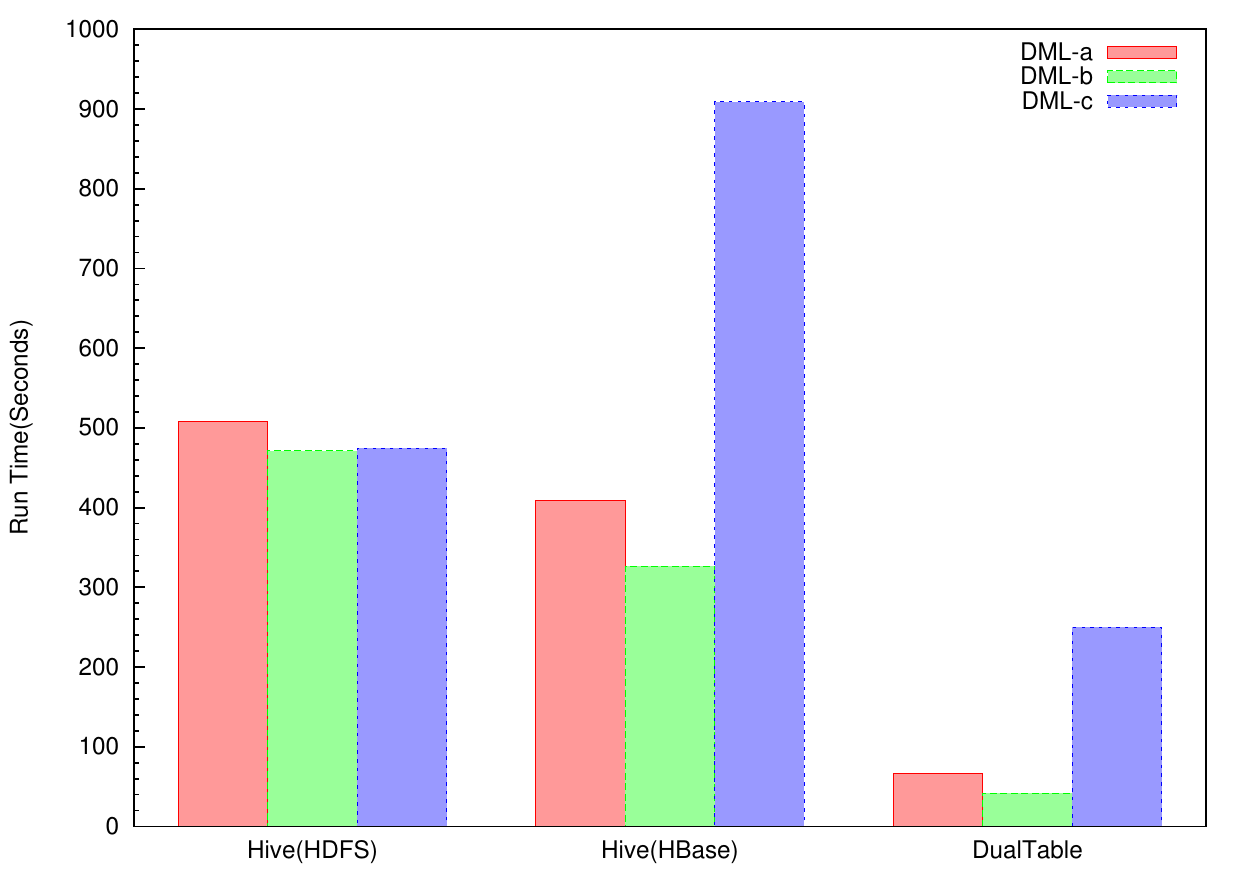}
 \caption{Update Performance on 30GB TPC-H Data Set}
 \label{fig:Chart2}
\end{minipage}%

\begin{minipage}[b]{0.3\textwidth}
 \centering
 \includegraphics[width=\columnwidth]{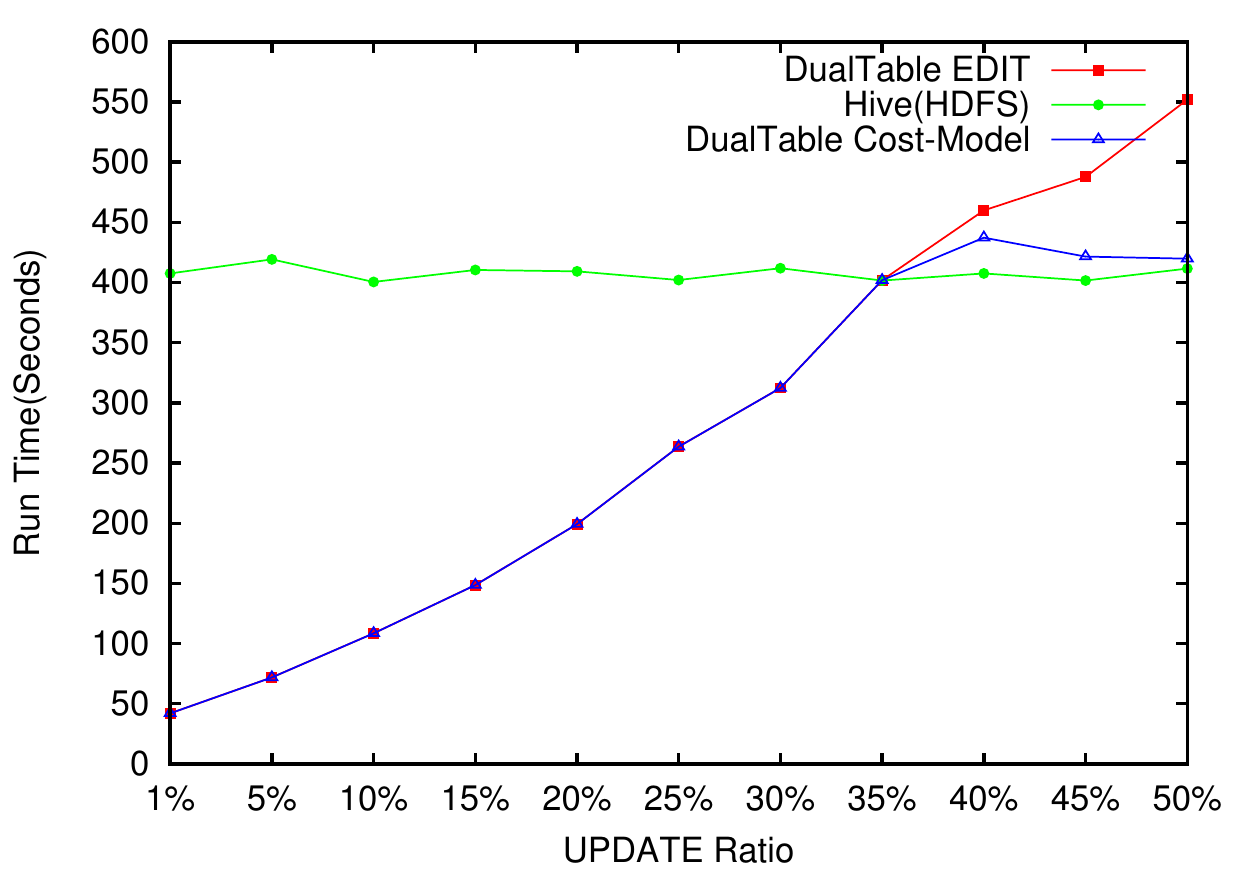}
 \caption{Update Performance for Different Workloads}
 \label{fig:Chart3}
\end{minipage}%
\hspace{0.05cm}
\begin{minipage}[b]{0.3\textwidth}
 \centering
 \includegraphics[width=\columnwidth]{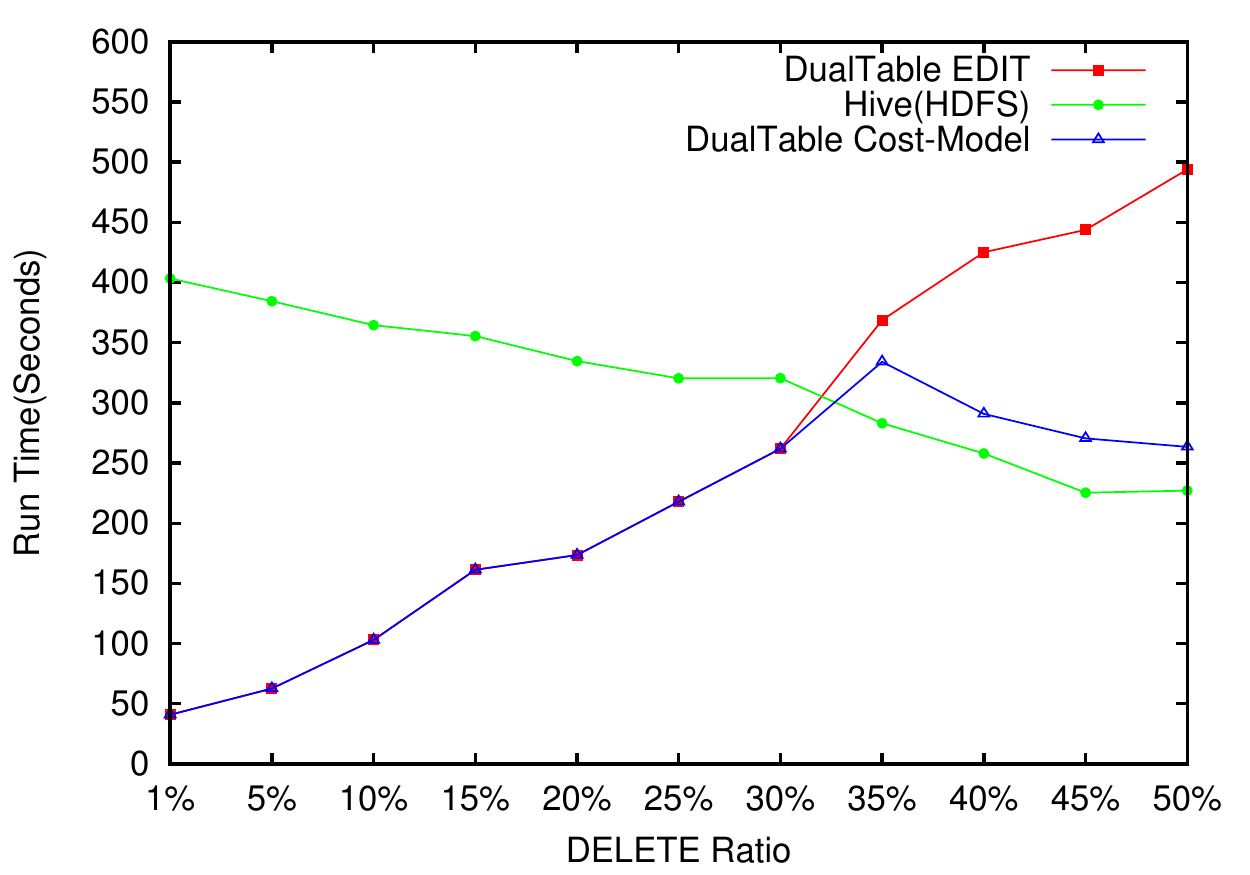}
 \caption{Delete Performance for Different Workloads}
 \label{fig:Chart4}
\end{minipage}
\hspace{0.05cm}
\begin{minipage}[b]{0.3\textwidth}
 \centering
 \includegraphics[width=\columnwidth]{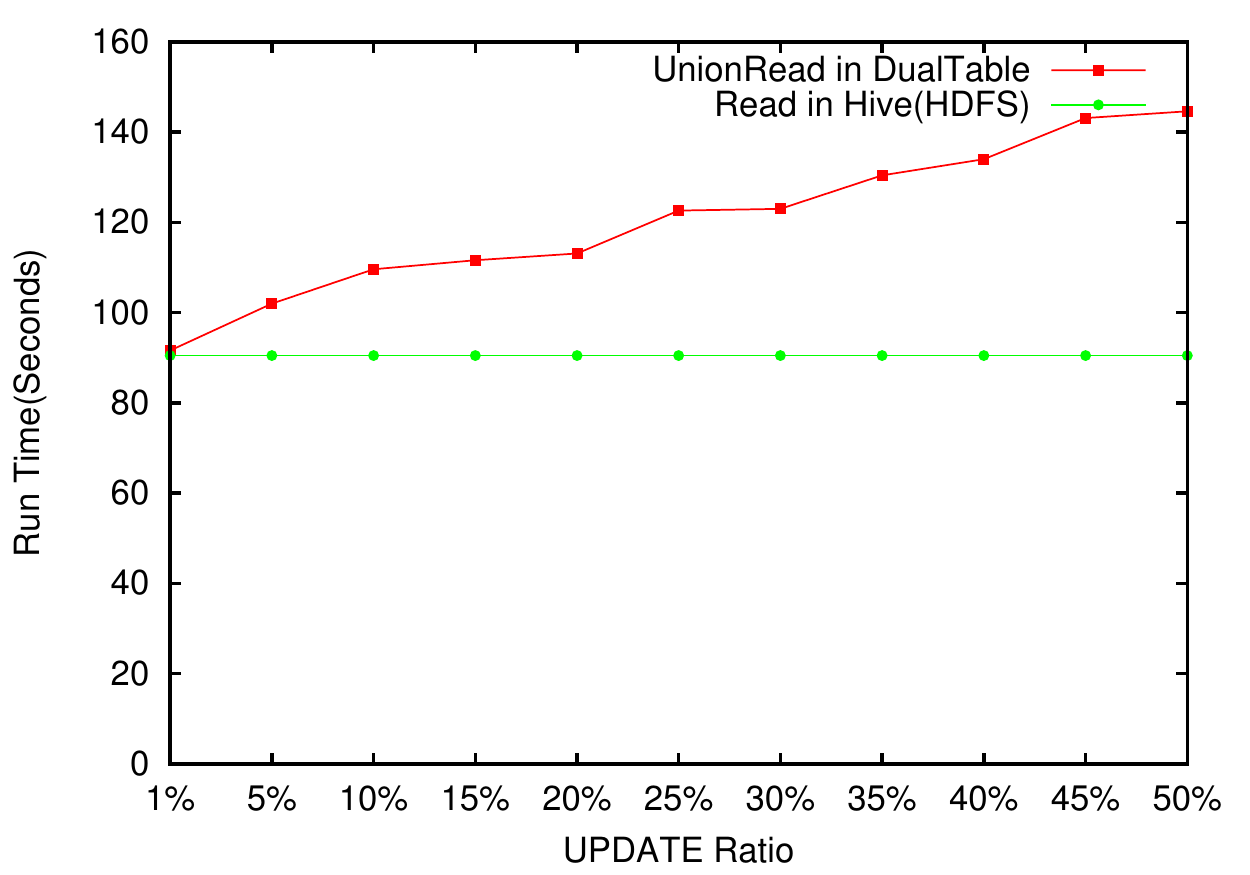}
 \caption{Overhead of Update Operations for Reads}
 \label{fig:Chart5}
\end{minipage}

\begin{minipage}[b]{0.3\textwidth}
 \centering
 \includegraphics[width=\columnwidth]{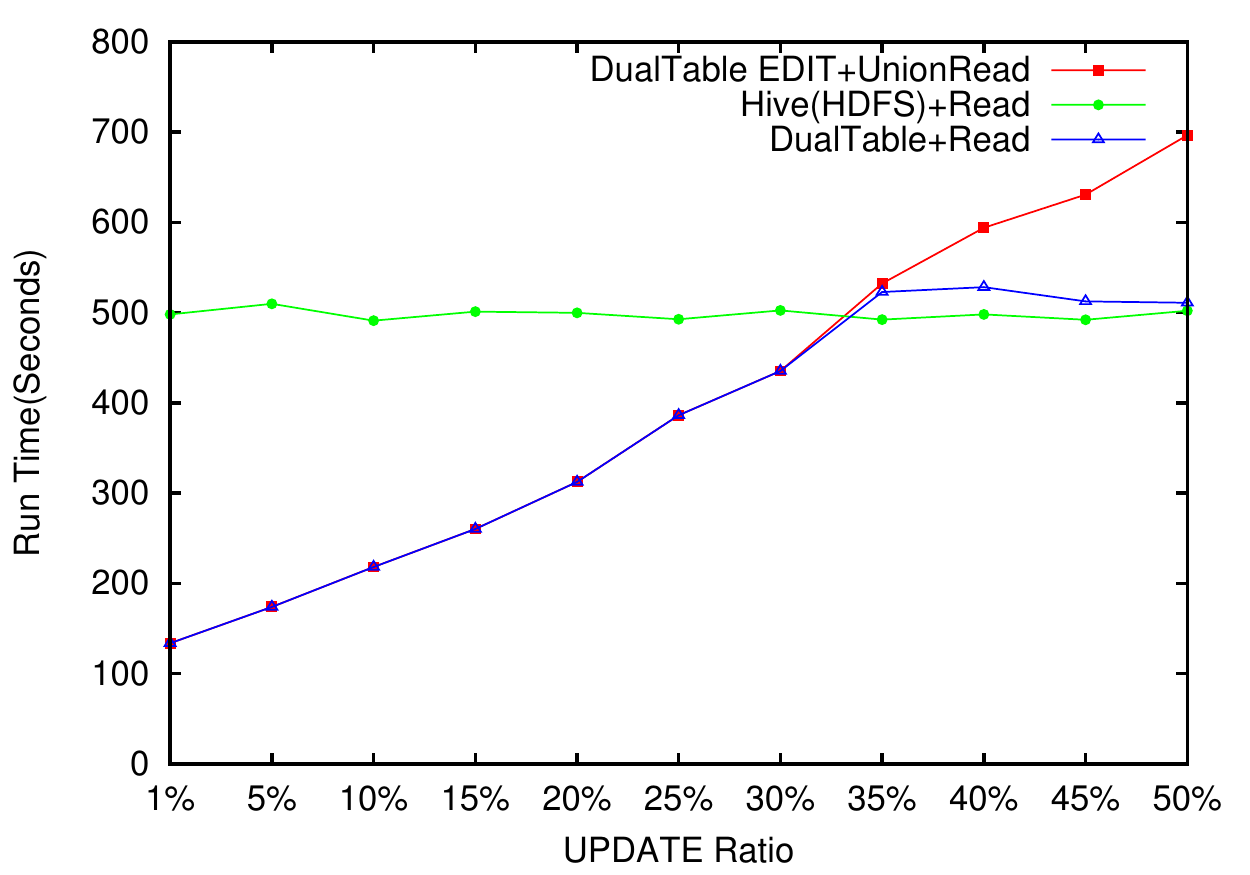}
 \caption{Update and Successive Read}
 \label{fig:Chart6}
\end{minipage}%
\hspace{0.05cm}
\begin{minipage}[b]{0.3\textwidth}
 \centering
 \includegraphics[width=\columnwidth]{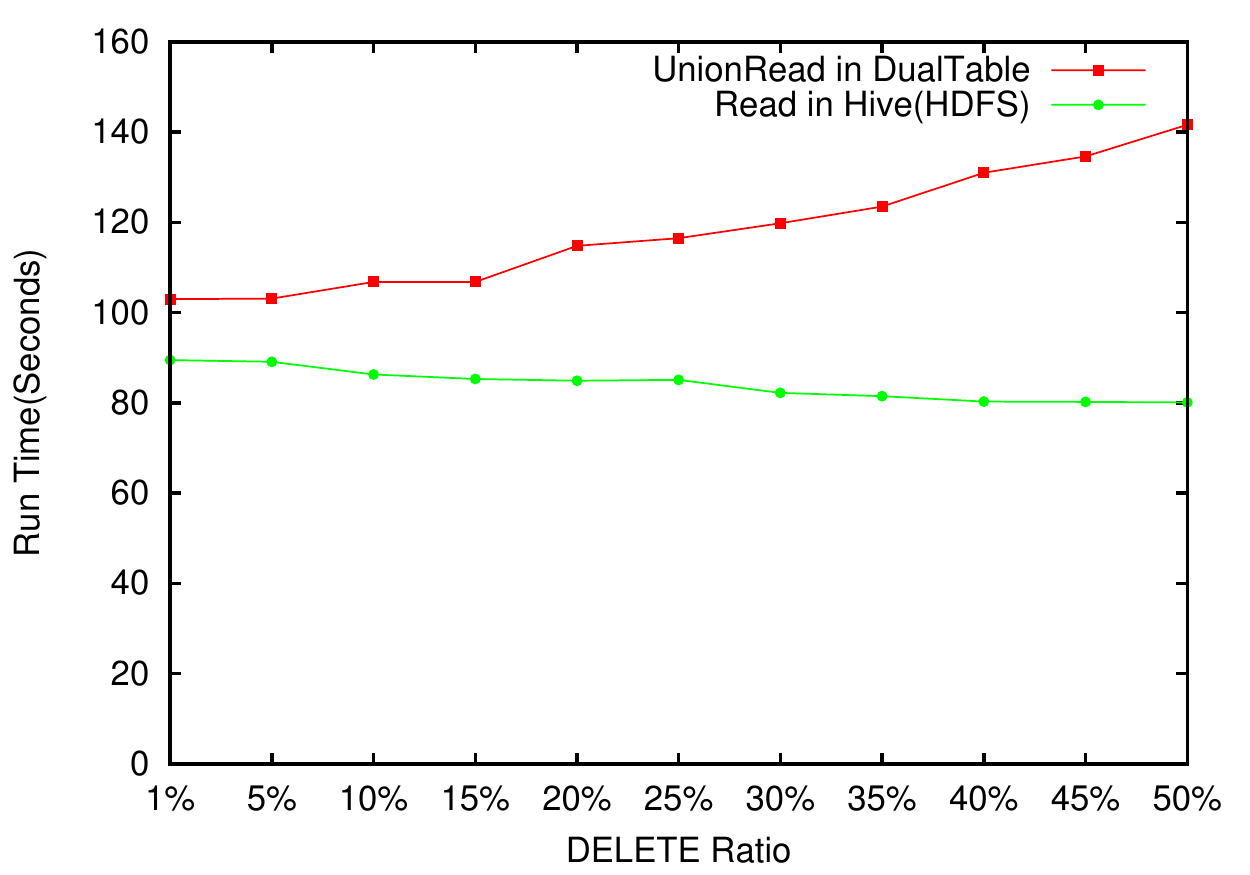}
 \caption{Overhead of Delete Operations for Reads}
 \label{fig:Chart7}
\end{minipage}%
\hspace{0.05cm}
\begin{minipage}[b]{0.3\textwidth}
 \centering
 \includegraphics[width=\columnwidth]{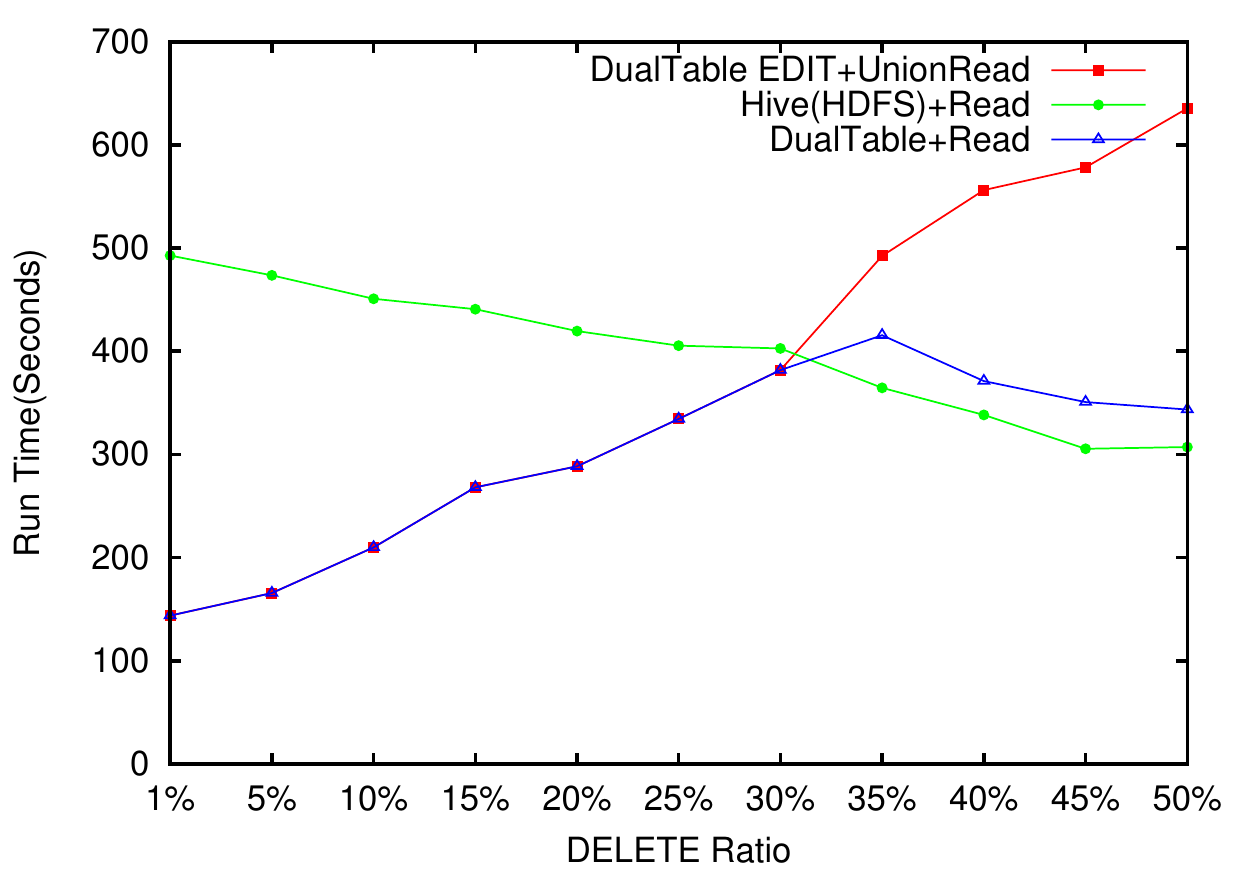}
 \caption{Delete and Successive Read}
 \label{fig:Chart8}
\end{minipage}

\end{center}
\end{figure*}

In the first experiment, we use 3 different queries to estimate the read efficiency of DualTable. Query a is
TPC-H query Q1, Query b is TPC-H Q12, and Query c is a count on the whole lineitem table. The Attached Table
is empty in this experiment. Thus, we measure DualTable's basic overhead, which is negligible as can be seen
in Figure \ref{fig:Chart1}.

In the second experiment, we run 3 typical update statements. DML-a updates 5\% of lineitem, DML-b deletes 2\% of
lineitem, and DML-c joins lineitem and order and updates 16\% of order. At the beginning of the experiment, the
Attached Table is empty. The performance results can be seen in Figure \ref{fig:Chart2}. As can be seen in the
figure, DualTable is most efficient for all updates, since it avoids unnecessary writes that Hive on HDFS
would have to perform, but features faster reads than HBase.

To assess the cost of DualTable's performance for different ratios of deletes and updates, we perform an additional
experiment. Starting with an empty Attached Table, we execute updates, which randomly update one field in
1\% to 50\% of the records in lineitem. In Figure \ref{fig:Chart3}, the performance for Hive, DualTable in EDIT mode, and
DualTable with the cost model can be seen. As expected, the performance of updates in Hive is constant for all
update ratios, while it increases with the amount of data changed in both versions of DualTable. The cross-over point is
reached at an update ratio of 35\%, when overwriting becomes cheaper than storing delta records in the Attached Table.
The cost model based DualTable changes to the OVERWRITE plan when the Attached Table becomes too costly
 and thus has a similar performance
to Hive from that point while the pure EDIT plan version gets more expensive.


%
%

In Figure \ref{fig:Chart4}, the same experiment is repeated for deletes. Unlike in the update case, the workload for
Hive becomes less with increasing delete ratio, since less data has to be written. Therefore, the cross-over point is
reached at a lower delete ratio. The delete cost model again finds the correct ratio to switch plans.



In Figure \ref{fig:Chart5}, the overhead of reading data from the Attached Table is shown. In the experiment,
we executed a full table scan after updating 1\% to 50\% of the lineitem table. While the read performance of
Hive is unaffected by the updates, since data is always rewritten, the DualTable UNION READ operation incurs
additional load to read data from both HDFS and HBase and merge it. The overhead in the update case is linear
to the amount of data in the Attached Table. In this experiment, no cost model was used. In Figure \ref{fig:Chart6},
the total cost of the update operation and an additional read are shown. This is the most realistic case,
where updates are performed and then the updated data set is analyzed. The results are similar to the pure
update experiments, with the difference that the cross over point is slightly below 35\% update ratio, which
is due to the additional overhead incurred for merging the data from the Master Table and the Attached Table
in the read query. The more often the data is read the lower the cross over point will be, which underlines the
importance of the cost model to ensure the best possible plan.



We repeated this experiment for the delete operation. The results can be seen in Figure \ref{fig:Chart7} and Figure \ref{fig:Chart8}.
The results confirm the results from previous experiments. Entries in the Attached Table incur an overhead for read operations, which
is more pronounced for high delete ratios since in Hive less data has to be read for the query part,
while DualTable keeps the original records and adds delete markers. Nevertheless, for delete ratios below
30\% DualTable is always more efficient than Hive. The cost model always chooses the best plan.

\section{Related Work} \label{sec:related}

Hive provides HiveQL, a declarative query language, which exposes an SQL-like
interface for Hadoop \cite{Thusoo2009}. Internally, Hive first translates HiveQL into a directed
acyclic graph (DAG) of MapReduce jobs and then executes the jobs in a MapReduce
environment.

From this point of view, there are three aspects or levels of optimization goals in Hive:
optimization of the query plan, especially when general SQL needs to be run in this environment;
optimization of the execution system, mostly including optimization of MapReduce
and development of compatible systems; and I/O optimization, which may include
optimized data placement, index creation, etc. Even though work in one aspect may
also involve contributions to some other aspects, related work can be categorized
into these three classes.

\subsection{Query Plan Optimization}
Hive itself only supports some basic rule-based optimization such as predicate push down and
multiple join strategies including MAP-join and Sort-Merge-Bucket join.

YSmart can detect correlated operations within a complex query, and use a
rule-based approach to simplify the whole query structure to generate a MapReduce
plan with minimal tasks \cite{Lee2011}. YSmart has been merged into
the official Hive version\footnote{\url{https://issues.apache.org/jira/browse/HIVE-2206}}. Sai Wu proposed a Hive optimizer called
AQUA \cite{Wu2011}, which can categorize join operations in one query into
several groups and choose the optimal execution plan of the groups based on a
predefined cost model. Xiaofei Zhang presented an approach to optimize multiple
path join operations in order to improve the overall
parallelization \cite{Zhang2012}. Harold Lim presented a MapReduce workflow optimizer called
Stubby, which uses a series of transformation rules to generate a set of
query plans and find the best one \cite{Lim2012}.
All of them attempt to solve the problem of translating SQL to MapReduce and
reorganizing the MapReduce DAG to yield better performance, focusing on
optimization at MapReduce level. Furthermore, QMapper considers variations of
SQL queries and their influences on query performance \cite{Xu:2013}. QMapper uses a
query rewrite-based approach to guide the translation procedure from SQL to a
variation of Hive queries and selects the best plan based on a modified cost model.
These works involve SQL-MapReduce or SQL-HiveQL-MapReduce translation, using
techniques like query graph analysis, query rewriting, and optimization of the DAG structure.
Their approach focuses on the MapReduce flow or higher layers and none of them
considers data manipulation within one MapReduce task. All of them choose to use
Hive-friendly storage, like HDFS, by default. Neither UPDATE nor DELETE operations
are discussed.


\subsection{Execution Environment Optimization}

To improve the performance or features of Hive, many HiveQL compatible systems have been developed, like
Shark \cite{Engle2012} based on Spark \cite{Zaharia2012},
Cloudera Impala \cite{Impala}, and others. Technologies for in-memory
processing, more efficient data reading and writing, and partial DAG execution
are utilized to enhance the whole system or just particular kinds of
applications like recursive data mining and ad-hoc queries.

Besides, by designing and analyzing MapReduce cost models, a large body of research has
been done to enable execution level optimization of MapReduce.
Starfish, as an example, focuses on automatic MapReduce job parameter configuration \cite{Herodotou2011}.
It makes use of a profiler to collect detailed statistics from
MapReduce executions and utilizes a what-if engine to stimulate the execution
and estimate the cost. An optimizer is utilized to minimize the cost of finding
a good configuration in a search space with combinatorial explosion.
The aforementioned Stubby also uses the what-if engine here to estimate cost for a MapReduce workflow \cite{Lim2012}.
MRShare aims at task sharing among queries that contain similar subtasks \cite{Nykiel2010}.
Optimal grouping of relevant queries based on the MapReduce cost model minimizes
redundant processing cost and improves the overall efficiency.



From the perspective of data manipulation, these works are similar to those of
query plan optimization.
They optimize MapReduce
tasks and plans, either through intelligent configuration of environment settings
or just by improving sharing among MapReduce tasks. Data manipulation operations
are out of their scope.

\subsection{I/O Optimization}

Optimized data placement is a common way to reduce data loading and reading cost.
The RCFile splits a data file into a set of row groups, each group
places data in a column-wise order \cite{He2011}.
With the help of the RCFile, a Hive application can efficiently locate its inputs
onto several data groups while avoiding reading
redundant columns of necessary rows.  RCFile and Hortonworks' ORC (Optimized RCFile)
are widely used in the Hive environment \cite{Leverenz2013}. Different from RCFile, LLama divides
data columns into groups, and provides another kind of data format, CFile, to
store them \cite{Lin:2011}. An index mechanism is used for efficient data look up.
It is shown that data loading and join performance can be improved.
Driven by the requirement of Smart Grid data process, we have also proposed DGFIndex, a new multiple range index technology \cite{Liu:2014},
which significantly improved the overall efficiency of multiple range query with a fair low cost of space occupation.

Similar to RDBMS, creating indexes can also be of great value for I/O performance
improvement. For now, Hive itself can support a compact index called
CIndex \cite{HIVE-417}. CIndex can enable
multi-dimensional queries at the cost of a large disk space for the index structure.
Hadoop++ also provides an index-structured file format to reduce the I/O cost during data processing \cite{Dittrich:2010}.

Data placement and index technologies try to minimize I/O to improve
the query performance, but they do not improve update operation support.
On the other hand, Hive indexes will result in additional cost of
reconstructing index structures for applications with update operations
implemented with INSERT OVERWRITE statement.

\section{Conclusion} \label{sec:conclusion}

In this paper, we have presented DualTable, a novel storage model for Hive.
DualTable stores data selectively in HDFS or in HBase. While new records are
always stored in HDFS, updates are either directly executed on HDFS or stored
as delta records in HBase. The storage location is dynamically chosen by
a cost model. Our experiments with standard industry benchmarks and real
data and workloads from the China State Grid show that DualTable outperforms
Hive by orders of magnitude in realistic settings.

In future work, we will evaluate other storage options for the Attached Table,
and compare the performance of DualTable with that of Hive ACID once it is available. Furthermore, we will investigate how the proposed storage model can be
incorporated in other big data analytic systems such as
Impala.
Additionally, we will investigate multi-query optimization in Hive, which
we expect to yield significant performance improvements in enterprise use cases such smart grid
data management.

\textbf{Acknowledgements}: This work is supported by the National Natural Science Foundation of China under Grant No.61070027, 61020106002, 61161160566.

\balance

\bibliographystyle{IEEEtran}
\bibliography{DualTable}
\end{document}